\documentclass[%
reprint, superscriptaddress, nobibnotes,
amsmath, amssymb, aps, prx
]{revtex4-2}

\usepackage{graphicx}
\usepackage{dcolumn}
\usepackage{bm}
\usepackage{float}
\usepackage{physics}
\usepackage{suffix}
\usepackage{xstring}
\usepackage{siunitx}
\usepackage{overpic}
\usepackage[english]{babel}

\usepackage{color}
\usepackage[normalem]{ulem} 
\definecolor{mgreen}{RGB}{1,123,0}

\usepackage{hyperref}
\hypersetup{
    colorlinks = true,
    linkcolor=blue,
    urlcolor=blue,
    citecolor=blue,
    linktoc=all, 
    bookmarksnumbered=true,
    pdfborder={0 0 0}, 
}
\makeatletter
\renewcommand{\fnum@figure}{FIG.\ \thefigure}
\makeatother

\begin{document}

\preprint{APS/123-QED}

\title{Coupling-induced universal dynamics in bilayer two-dimensional Bose gases}

\author{En Chang}
\email{en.chang@physics.ox.ac.uk}
\affiliation{Clarendon Laboratory, University of Oxford, Parks Road, OX1 3PU, United Kingdom}

\author{Vijay Pal Singh}
\affiliation{Quantum Research Centre, Technology Innovation Institute, Abu Dhabi, UAE}

\author{Abel Beregi}
\affiliation{Clarendon Laboratory, University of Oxford, Parks Road, OX1 3PU, United Kingdom}

\author{Erik Rydow}
\affiliation{Clarendon Laboratory, University of Oxford, Parks Road, OX1 3PU, United Kingdom}

\author{Ludwig Mathey}
\affiliation{Zentrum f\"ur Optische Quantentechnologien and Institut f\"ur Quantenphysik, Universit\"at Hamburg, 22761 Hamburg, Germany}
\affiliation{The Hamburg Centre for Ultrafast Imaging, Luruper Chaussee 149, Hamburg 22761, Germany}

\author{Christopher J. Foot}
\affiliation{Clarendon Laboratory, University of Oxford, Parks Road, OX1 3PU, United Kingdom}

\author{Shinichi Sunami}
\email{shinichi.sunami@physics.ox.ac.uk}
\affiliation{Clarendon Laboratory, University of Oxford, Parks Road, OX1 3PU, United Kingdom}

\date{\today}

\begin{abstract}
The emergence of order in many-body systems and the associated self-similar dynamics governed by dynamical scaling laws is a hallmark of universality far from equilibrium.
Measuring and classifying such nontrivial behavior for novel symmetry classes remains challenging.
Here, we realize a well-controlled interlayer coupling quench in a tunable bilayer two-dimensional Bose gas, driving the system to an ordered phase.
We observe robust self-similar dynamics and a universal critical exponent consistent with diffusion-like coarsening, driven by vortex and antivortex annihilation induced by the interlayer coupling. 
Our results extend the understanding of universal dynamics in many-body systems and provide a robust foundation for quantitative tests of nonequilibrium effective field theories.
\end{abstract}	
\maketitle

\section{Introduction}
Ordering dynamics far from equilibrium is among the most intriguing areas in many-body physics, with implications extending from early-universe cosmology to condensed matter and particle physics~\cite{ micha_Relativistic_2003,kibble_topology_1976,zurek_cosmological_1985,chuang_cosmology_1991,berges_turbulent_2014}. Systems driven far from equilibrium have been shown to exhibit scale invariance and self-similar behavior as they enter universal scaling regimes defined by specific scaling laws and exponents~\cite{bray_theory_1994, prufer_observation_2018, mikheev_universal_2023, huh_universality_2024}.
Such universal dynamics have been reported across various systems spanning spinor Bose condensates~\cite{lamacraft_quantum_2007,williamson_universal_2016,bourges_different_2017,mukerjee_dynamical_2007,schmied_violation_2019,williamson_coarsening_2017,huh_universality_2024}, binary mixtures~\cite{hofmann_coarsening_2014,fujimoto_scaleinvariant_2020,singh_thermal_2023}, spin-orbit-coupled condensates~\cite{rajat_collective_2025}, exciton-polariton condensates~\cite{kulczykowski_phase_2017,comaron_dynamical_2018,gladilin_multivortex_2019}, Rydberg atom arrays~\cite{manovitz_quantum_2025} and vortex-free scalar Bose gases~\cite{gazo_universal_2025}.
A particularly interesting form of universal dynamics is coarsening, which emerges after a quench from disorder to order when the domain size grows beyond the relevant microscopic scale. It is well established that coarsening can follow a spontaneous symmetry breaking transition~\cite{bray_theory_1994,williamson_universal_2016, williamson_coarsening_2017}, or arise from topological transition, as in the 2D XY model where vortex and antivortex annihilation drive ordering~\cite{karl_strongly_2017,comaron_quench_2019,groszek_crossover_2021}. In both cases, the late-time growth typically follows power laws set by dimensionality, symmetry, conservation laws, and topological defect types, rather than by microscopic details~\cite{bray_theory_1994,huh_universality_2024}, underscoring universality out of equilibrium. 
However, whether coarsening dynamics can arise following a quench that explicitly breaks the original symmetry remains poorly understood.

\begin{figure*}[t]
    \centering
     \includegraphics[width=\linewidth]{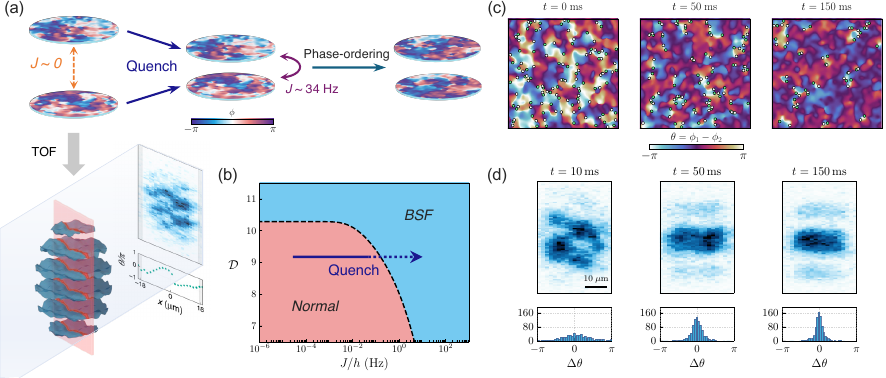} 
    \caption{Interlayer coupling quench in bilayer 2D quantum gases.
    (a) (Top) 
     We quench the interlayer coupling $J$ from nearly zero to $34$ Hz, 
    driving an initially decoupled bilayer with free vortices toward a phase-locked, coherent state. 
    (Bottom) Relaxation dynamics are monitored by matter-wave interferometry, with local fluctuations of relative phase captured by optically pumping a thin slice (red sheet) before absorption imaging. A representative pre-quench interference image is shown, with the extracted relative phase profile plotted below.
    (b)
    The equilibrium phase diagram for the relative-phase mode, as a function of the interlayer coupling $J$ and phase-space density $\mathcal{D}$, obtained via renormalisation-group analysis~\cite{mathey_phaselocking_2007,rydow_observation_2025}; the quench induced by the sudden increase in $J$ drives the system to a bilayer superfluid (BSF). The dashed line indicates the critical points.
    (c) 
    Numerical simulation of the quench dynamics shows relaxation toward a nearly phase-locked state. 
    Vortices (open circles) and antivortices (filled circles) decay over time through dynamical pairing and annihilation. 
    The phase-locked domains (dark blue to dark red) grow progressively.
    (d)
    (Top) Typical single-shot interference image at different hold times after the quench show the evolution from an initial state containing vortices (left), identifiable by sharp phase discontinuities (see Fig.~\ref{fig:corr_vortex_analysis}), to a phase-coherent state (right). (Bottom) Histogram of phase difference $\Delta\theta=\theta(x)-\theta(x^\prime)$ at fixed distance $\left|x-x^{\prime}\right|=\SI{5}{\micro \meter}$, showing suppression of phase fluctuations over time, obtained from 40 experimental runs. }
 \label{fig:concept}
\end{figure*}

The coupled 2D XY model is a paradigmatic platform for such an investigation, with broad relevance including high temperature superconductivity~\cite{zhang_vortices_2005,leggett_what_2006,benfatto_kosterlitzthouless_2007,okamoto2017,homann_dissipationless_2024} and high energy physics~\cite{Eto2018,Kobayashi2019,Tylutki2016}. In equilibrium, the relative mode of the system exhibits a normal to coupling-induced bilayer superfluid (BSF) transition driven by vortex pair-binding, with a critical point that has a strong dependence on the interlayer coherent coupling strength $J$, which explicitly breaks the $U(1)$ symmetry~\cite{cazalilla_competition_2007,benfatto_kosterlitzthouless_2007,mathey_phaselocking_2007,bighin_berezinskiikosterlitzthouless_2019,xiao_fate_2025}. 
Therefore, the recent realization of highly tunable coupled bilayer superfluids with ultracold atoms~\cite{rydow_observation_2025} offers a unique opportunity to extend the experimental probe of nonequilibrium dynamics: the precise and rapid control of the interlayer coupling strength in such a setup enables sudden coupling quenches. 
In the absence of topological defects, theory predicts that the relative phase following a coupling quench undergoes universal rephasing connected to sine-Gordon (SG) dynamics in both 1D and 2D superfluids~\cite{dallatorre_universal_2013,schweigler_experimental_2017,horvath_nonequilibrium_2019,tononi_dephasing_2020,bastianello_sinegordon_2024,szasz-schagrin_nonequilibrium_2024}. 
In contrast, the coupled 2D XY model hosts vortices and antivortices in the relative mode whose presence qualitatively modify relaxation and can enable a transition from disorder to order, leaving open whether universal dynamics persist and, if so, what form they take.

In this work, we report experimental observations of universal dynamics following a symmetry breaking coupling quench, appearing as coarsening via vortex and antivortex annihilation in the relative phase of a coherently coupled 2D bilayer.
To this end, we realize a bilayer 2D quantum system consisting of two nearly homogeneous layers of 2D Bose gases with precisely and dynamically controlled interlayer coupling. 
This allows us to quench the system from a disordered phase to an ordered phase by a sudden change of the coupling, after which the system relaxes toward a phase-locked, coherent state in the relative phase mode. 
We probe the resulting relaxation dynamics using matter-wave interferometry, 
which provides direct access to local phase fluctuations in the system. 
From these, we measure the two-point correlation function and the vortex density, which are key observables for characterizing the scale-invariant universal dynamics. 
The typical size of the phase coherent domain, extracted from the correlation function, exhibits power-law scaling in direct agreement with the scaling of the vortex density. 
By varying the initial phase-space density (PSD), we verify the universality of phase-ordering dynamics and extract the corresponding dynamic critical exponent $z = 1.73(9)$.  
We perform extensive numerical simulations based on classical-field approximation to corroborate our measurements. 
The measured value of the critical exponent indicates a diffusion-like coarsening, consistent with theoretical predictions for scaling dynamics in quenched 2D Bose gases~\cite{groszek_crossover_2021}, and close to the (near-) Gaussian non-thermal fixed point (NTFP) predicted in both quenched 2D Bose gases and the 2D SG model~\cite{karl_strongly_2017,heinen_anomalous_2023}.

\section{Coupling quench in bilayer 2D Bose gases }

We perform experiments with Bose gases of $^{87}$Rb atoms trapped in multiple RF(MRF)-dressed bilayer potentials, as detailed in Refs.~\cite{barker_coherent_2020,sunami_observation_2022,beregi_quantum_2024}.
Combined with a ring-shaped optical dipole potential, we realize a nearly homogeneous bilayer 2D Bose gas~\cite{sunami_detecting_2025, rydow_observation_2025} with precise control of the coupling between the layers. 
For two tunnel-coupled 2D Bose gases labelled $i=1,2$ and each represented by bosonic field $\Psi_i(\boldsymbol{r})$, it is convenient to decompose the system into symmetric (common) and anti-symmetric (relative) degrees of freedom. 
Common and relative phase modes are of particular interest, defined respectively as $\varphi(\boldsymbol{r})=\left(\phi_1(\boldsymbol{r})+\phi_2(\boldsymbol{r})\right)/2$ and $\theta(\boldsymbol{r})=\phi_1(\boldsymbol{r})-\phi_2(\boldsymbol{r})$~\cite{whitlock_relative_2003,grisins_coherence_2013}, where $\phi_i = \arg(\Psi_i)$ are the phase modes of each layer~\cite{rydow_observation_2025}. 
In the absence of interlayer coupling, the relative and common phases undergo the well-known Berezinskii–Kosterlitz–Thouless (BKT) transition. With coupling, phase locking is energetically favored, establishing coherence between the two layers and driving the system into the BSF phase~\cite{mathey_phaselocking_2007, rydow_observation_2025}. 
The coupling term explicitly breaks the $U(1)$ symmetry of the relative phase, resulting in a distinct behavior from the conventional BKT transition. However, the transition in relative phase is still expected to originate from the binding of vortex-antivortex pairs~\cite{fertig_deconfinement_2002,fertig_vortex_2003,zhang_vortices_2005}. 

Our experiments start with a cloud of atoms in equilibrium, with a total atom number ranging from $1.8 \times 10^4$ to $2.5 \times 10^4$ at $T=\SI{27}{\nano \kelvin}$, in a decoupled double well with a barrier height of $\SI{8.7}{\kilo \hertz}$ and a well separation of $\SI{3.25}{\micro \metre}$~\cite{sunami_observation_2022, rydow_observation_2025}. 
The barrier height is then rapidly lowered to $\SI{1.9}{\kilo \hertz}$ over $\SI{15}{\milli \second}$, reducing the well separation to $\SI{2.18}{\micro \metre}$ and increasing the coupling strength $J/h$ from nearly $\SI{0}{\hertz}$ to $\SI{34}{\hertz}$. 
For the initial trap configuration, the coupling is estimated to be less than $10^{-3}$ Hz, which we refer to as zero coupling~\cite{rydow_observation_2025}.
The 15 ms quench induces a sudden change in the characteristics of the 2D system while
being adiabatic for the tightly confined vertical degrees of freedom thus realizes a quench scenario in which the undesired excitations are suppressed.

\begin{figure}[t]
    \centering
     \includegraphics[width=\linewidth]{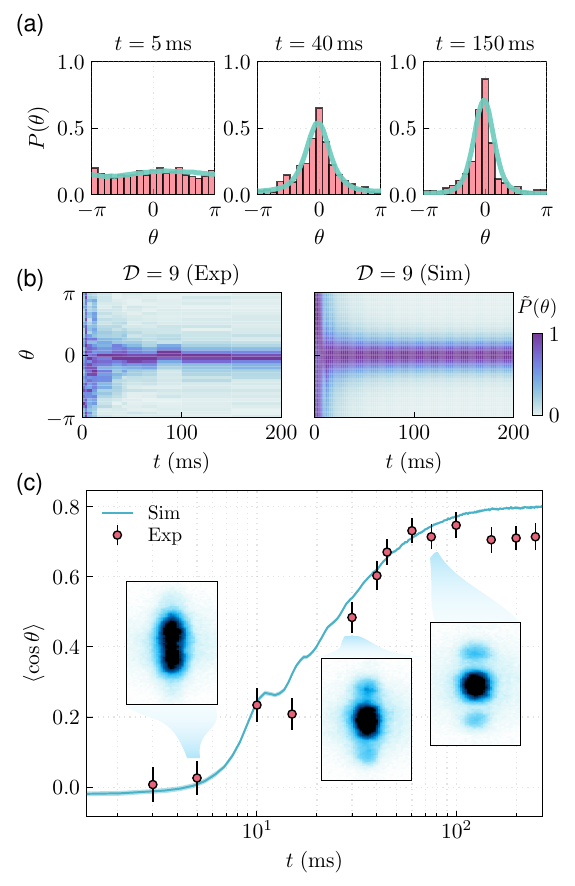}
    \caption{Dynamical phase-locking transition.
    (a) Probability density distribution of the relative phase, $P(\theta)$, at selected evolution times after the quench, obtained from experiments (histograms) and simulations (solid lines).
    (b) Time evolution of the normalized phase distribution, $\tilde{P}(\theta) = P(\theta)/\max(P(\theta))$, from experiments (left panel) and simulation (right panel), illustrating the emergence of phase locking.
    Each column represents a histogram of the phases (panel a) by colors.
    (c) Relative-phase order parameter $\langle\cos\theta\rangle$ plotted on a linear-log scale. 
    Experimental data (dots) have error bars obtained via bootstrapping, while simulation results are shown as a solid line (mean) with a shaded region indicating uncertainty. The simulation exhibits a small, damped oscillation at a frequency on the order of the Josephson plasma frequency. Averaged interference patterns from over 40 experimental runs are shown for selected evolution times $t$ (insets).
    }
     \label{fig:locking}
\end{figure}

Throughout the experiment, the quasi-2D conditions $\hbar \omega_z > k_B T$ and $\hbar \omega_z > \mu$ are satisfied, where $\hbar$ is the reduced Planck constant, $k_B$ the Boltzmann constant, and $\mu$ is the chemical potential, and the axial trapping frequencies are $\omega_z/2\pi = \SI{1.2}{\kilo \hertz}$ for both layers after the quench. 
The characteristic dimensionless 2D interaction strength is $\tilde{g} = \sqrt{8\pi} a_s/\ell_0\sim0.08$, where $a_s$ is the 3D scattering length and $\ell_0=\sqrt{\hbar/(m\omega_z)}$ is the harmonic oscillator length along $z$ for an atom of mass $m$.
The quench drives the system far from equilibrium, where the final coupling strength corresponds to the BSF phase (Fig.~\ref{fig:concept}(b)), such that
the system dynamically phase-locks and relaxes to an ordered state. To study the relaxation dynamics, atoms are held in the double well for various durations $t$ after the quench, then released for time-of-flight (TOF) expansion ($t_{\text{TOF}}=\SI{16.5}{\milli \second}$).  
The interference pattern from the two clouds during TOF encodes the relative phase of the system; this is detected in a spatially selective manner, such that atoms within a thin slice ($L_z = \SI{4}{\micro \metre}$) are probed by absorption imaging (Fig.~\ref{fig:concept}(a)). 
This allows us to obtain the local relative phases of the system~\cite{sunami_observation_2022}. We ensure balanced atom populations in the two layers by maximizing the observed matter-wave interference contrast~\cite{barker_coherent_2020,sunami_detecting_2025}, and repeat the measurements at least 40 times at each atom number and hold time $t$, to obtain sufficient statistics for the analysis we describe below.

\section{Universal phase ordering dynamics}

\begin{figure}[tb]
    \centering
     \includegraphics[width = \linewidth]{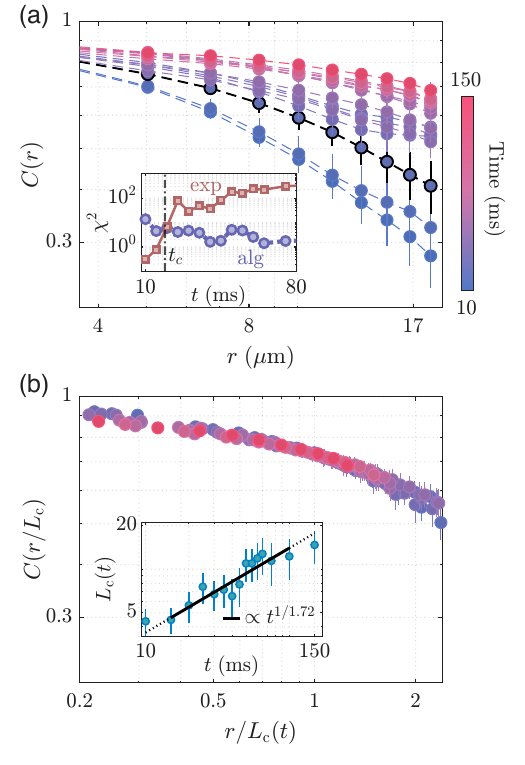} 
    \caption{Build-up of phase coherence and the self-similar dynamics. 
    (a) Phase correlation function $C(r, t)$ at various hold times after the quench, averaged over 50 realizations with error bars showing the standard error. 
     The spatial decay is fitted with both algebraic and exponential decay models, revealing a transition to algebraic decay at a critical time $t_c$, as indicated by the crossover of reduced $\chi^2$ statistics of the fits (inset). 
     The correlation function at $t_c$ is highlighted by the circular markers with a black outline, connected by black dashed line. 
     We define a characteristic correlation length $L_c(t)$, serving as a proxy for the domain size, by the condition  $C(L_c(t), t) = 0.75$; see text for details. 
    (b) Correlation functions at different times beyond $t_c$ collapse onto a single curve, when plotted on a log-log scale.
    Inset:  Power-law growth of the domain size $L_c(t)$, shown on a log-log plot. 
    The black line is a fit to the form $L(t) = A\times t^{1/z_\mathrm{corr}}$.
    The solid portion of the line denotes the scaling window used for the fit. Error bars denote standard error.
     }
 \label{fig:corrcollapse}
\end{figure}

Following the quench, the interlayer coupling begins to lock the phases of the two layers. 
This corresponds to the emergence of a preferred relative phase that minimizes the $-\cos(\theta)$ term in the effective Hamiltonian (see Eq.~\ref{eq:Hamiltonian} in~Appendix~\ref{app:effective_low_energy}), reflecting the breaking of $U(1)$ symmetry with minima at $\theta=2\pi n$, $n\in \mathbb{Z}$. In Fig.~\ref{fig:locking}, we show the distribution of the relative phase extracted from the matter-wave interference pattern at various hold times. The distribution is nearly uniform shortly after the quench, consistent with the initial decoupled state possessing $U(1)$ symmetry. The onset of phase locking is evidenced by the gradual development of a peak at wrapped phase $\theta=0$ over time. 
To quantify the phase locking, we plot the evolution of the phase order parameter $\langle\cos\theta\rangle$, which is indicative of phase-locking strength and interference contrast~\cite{dallatorre_universal_2013,schweigler_experimental_2017}. 
The averaged experimental interference fringes exhibit a clear increase in interference contrast over time, and the resulting time evolution of $\langle\cos\theta\rangle$ is in good agreement with numerical simulations that account for finite imaging resolution. The saturation of $\langle\cos\theta\rangle$ at slightly lower values than the numerical simulation arises from finite thermal populations that become more pronounced at longer evolution times.

Fig.~\ref{fig:corrcollapse} shows the equal-time phase correlation function $C(r) = C(\overline{x}) = \Re \left[ \langle e^{i[\theta(x)-\theta(x-\overline{x})]} \rangle_x \right]$ at different times after the quench.
We observe an increase in phase coherence across the system following the quench, demonstrated by the growth of the spatial correlation functions.
In Fig.~\ref{fig:corrcollapse}(a), we observe that the decay function changes from exponential $C(r) \sim e^{-2r/r_0}$, with correlation length $r_0$, to power law $C(r) \sim r^{-2\eta}$, with exponent $\eta$, 
as verified by the $\chi^2$ test, reflecting the gradual phase ordering from the disordered (normal) phase. 

\begin{figure*}[t!]
    \centering
     \includegraphics[width=\linewidth]{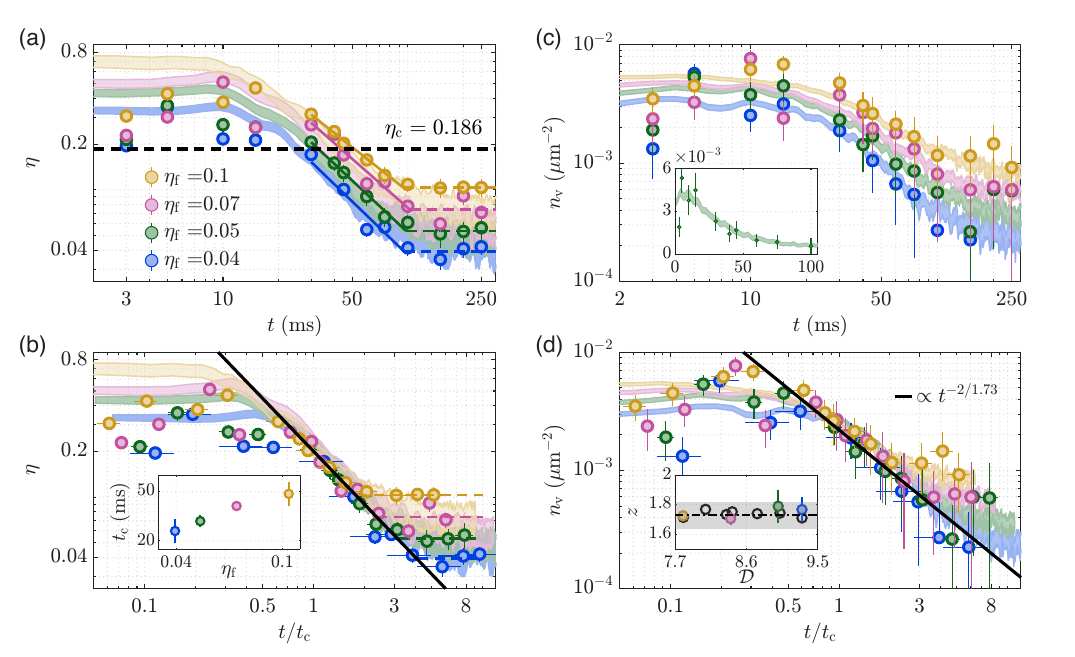}
 \caption{Universal scaling behavior. 
 (a) Temporal evolution of $\eta(t)$, determined by fitting $C(r)$ with an algebraic model, for four different initial conditions, plotted on a log-log scale. Measurements (dots with error bars denoting fit uncertainties) are shown alongside simulations (represented by shaded regions indicating uncertainty). 
 The horizontal dashed line marks the critical value $\eta_c$ for the onset of algebraic phase coherence. The solid lines are a guide to the eye, illustrating that $\eta$ follows a power-law behavior in the scaling window. 
 (b) Rescaled time evolution of $\eta(t/t_{c})$, demonstrating universal dynamics (indicated by the black solid line). The horizontal error bars arise from the uncertainty in $t_{c}$.
 The inset shows the critical time $t_{c}$ for different initial conditions.
 (c) Time evolution of the vortex density $n_{v}(t)$ for different initial conditions on a log-log scale, with error bars indicating the standard error.
 For $\eta_\mathrm{f}= 0.04$ and $t > 150$ ms, no vortices were observed in the finite dataset obtained in this work.
 The inset is linear plot for a selected initial condition.
 (d) Rescaled vortex density $n_{v}(t/t_{c})$, showing universal dynamics. The solid line indicates a power-law behavior consistent with the scaling $t^{-2/1.73}$. 
 The inset shows the fitted values of the dynamic critical exponent for different initial conditions; the dashed line indicates the mean value, and the shaded area represents the associated uncertainty. Open circles denote the exponent values extracted from simulations using the same fitting method.}
 \label{fig:eta_nv}
\end{figure*}

According to the dynamical scaling hypothesis, the correlation functions in the scaling region should evolve in terms of the dynamical scaling form $C(r, t) \sim F\left(r/L_c(t)\right)$ (see Fig.~\ref{fig:universal_correlation_scaling} in Appendix~\ref{app:univresal_corr}), and the growth of the correlation length (domain size) follows $L_c(t) \sim t^{1/z}$~\cite{bray_theory_1994}. We determine the characteristic correlation length scale by the distance at which the correlation function satisfies $C(L_c(t), t) = \text{const.}$~\cite{jelic_quench_2011,nam_coarsening_2012,williamson_universal_2016,kulczykowski_phase_2017,comaron_quench_2019,groszek_crossover_2021}; we select this constant value to be centered at 0.75, with resampling in the region $0.7 - 0.8$ to assess the robustness of the obtained $L_c$ values. 
This methodology gives robust results whilst allowing us to compute the mean and standard deviation of $L_c$. 
The correlation functions at various times after $t_c$ collapse onto a common curve when rescaled as $r' = r / L_c(t)$, indicating self-similar dynamics (Fig.~\ref{fig:corrcollapse}(b)). The extracted $L_c$ exhibits power-law scaling in the scaling window between $t\approx t_c$ and the time at which the system reaches steady state (see Fig.~\ref{fig:eta_nv}), with a critical exponent $z_\mathrm{corr} \approx 1.72$. 
This value is consistent with the exponent obtained from vortex-scaling analyses across different initial conditions, as discussed below.

Having verified the scale-invariant dynamics, we now analyze its universal characteristics by varying the initial PSD.
In Fig.~\ref{fig:eta_nv}(a), we plot $\eta (t)$ for various initial PSDs ($\mathcal{D} = 7.8$ to $\mathcal{D} = 9.3$). 
The values of $\eta (t)$ at the critical time $t_c$, at which the correlation function becomes better described by a power-law decay, give the critical exponent $\eta_{c}=0.186(4)$ at the onset of the phase-locking transition, which is nearly independent of the initial conditions.
At short timescales ($t < \SI{10}{ms}$), we observe an increase in $\eta$, indicating that the system becomes more disordered. This arises from the spontaneous generation of vortices following a rapid quench, as evidenced by the initial increase in vortex number shown in Fig.~\ref{fig:eta_nv}(c). Subsequently, fluctuations in $\eta(t)$ suggest that the system exhibits a delayed response to the phase transition. Around $t = 30$ ms, the system begins to phase-order, and $\eta(t)$ follows a power-law decay in the scaling region, as indicated by the solid lines in Fig.~\ref{fig:eta_nv}(a). After $\sim \SI{100}{ms}$, $\eta(t)$ reaches a constant value, indicating that the relative mode has attained its steady state. 
After rescaling the hold time to $t' = t/t_{c}$, we find that $\eta(t)$ begins to exhibit a universal decay near the critical time $t/t_{c}=1$, following a power-law scaling, as shown in Fig.~\ref{fig:eta_nv}(b).

To demonstrate that the phase-ordering process is governed by the annihilation of topological defects, we analyze the time evolution of the vortex density for the same datasets in Fig.~\ref{fig:eta_nv}(c). We observe power-lay decay of the vortex density $ n_v(t)$, indicative of dynamical vortex–antivortex pairing and annihilation. 
Concurrently, based on the dynamical scaling hypothesis, universal behavior should emerge around and after the critical time $t_{c}$, when the correlation length $L_{c}$ becomes the dominant length scale. The growth of the correlation length $L_{c}(t)$ and the reduction in vortices $n_{v}(t)$ are consistent within this framework, predicting that 
$ n_v(t)$ should scale as $n_v(t) \sim L_c^{-2}(t) \sim t^{-2/z}$~\cite{comaron_quench_2019,groszek_crossover_2021}. After rescaling the hold time to $t^{\prime}=t/t_{c}$ (Fig.~\ref{fig:eta_nv}(d)), we observe the universal decay of vortex density following the scaling law $n_v(t) \sim t^{-2/z}$ as predicted by theory, which indicates the system coarsens. However, we find that the decay of free vortices does not continue indefinitely, reaching a plateau after $\approx 100$ ms, consistent with the evolution of $\eta$. 

To determine the value of $z$, each vortex‐decay dataset is first fitted individually using the model $f(t) = A \times t^{-2/z}$, where $A$ and $z$ are free parameters (Fig.~\ref{fig:eta_nv}(d) inset). 
The fitting procedure is restricted to data points within the scaling window, defined as those that collapse onto a common decay, {0.6 $< t/t_c<$ 3}. To boost the bootstrap sample size, we first rescale all time axes to collapse the data onto a single decay curve. We then apply bootstrap fitting to the pooled, collapsed data, which yields a mean exponent $\langle z \rangle = 1.73$, with a standard error of 0.09 (see Fig.~\ref{fig:bootstrap_fitting}). It is close to the numerical results for the quenched 2D Bose gas ($z \approx 1.74$ is reported in both the conservative limit and under weak dissipation in Ref.~\cite{groszek_crossover_2021}. Ref.~\cite{karl_strongly_2017} finds $\beta \approx 0.56$ for near-Gaussian NTFP, corresponding to $z = 1/\beta \approx 1.79$), suggesting that the coarsening dynamics in the coupled bilayer follow the diffusion-type scaling, characterized by the dynamic critical exponent $z \sim 2$.

\section{Conclusion and outlook}
Our work demonstrates universal phase-ordering dynamics in a coupled two-dimensional bilayer system following a coupling quench, revealing good agreement between experimental observations and classical-field simulations. 
We observe self-similar dynamics in which the correlation length exhibits power-law scaling. This coarsening proceeds via vortex–antivortex annihilation, and the vortex density decays according to diffusion-like scaling $n_v(t) \sim t^{-2/z}$ with a dynamic critical exponent $z = 1.73(9)$.
Our results advance the understanding of universal dynamics following an explicit symmetry breaking quench and provide a robust foundation for testing effective field theories of out-of-equilibrium systems, such as the coupled 2D XY model and 2D sine-Gordon model with defect excitations.
Furthermore, the coupling quench serves as a novel and highly tunable approach to exploring a wide range of nonequilibrium dynamics in 2D quantum systems, including the Kibble-Zurek (KZ) mechanism~\cite{kibble_topology_1976,zurek_cosmological_1985,mathey_phaselocking_2007}, light-cone dynamics~\cite{mathey_light_2010}, reverse-KZ mechanism~\cite{mathey_light_2010,sunami_universal_2023}, universal rephasing~\cite{dallatorre_universal_2013}, and Josephson effects ~\cite{berrada_integrated_2013,spagnolli_crossing_2017,levy_ac_2007,pigneur_relaxation_2018,luick_ideal_2020}, many of which are inaccessible with the conventional single-layer approach.

\maketitle
\section*{Acknowledgements}
This work was supported by the EPSRC Grant Reference EP/X024601/1. A.B. and E.R. thank the EPSRC for doctoral training funding.  
L.M. acknowledges funding by the Deutsche Forschungsgemeinschaft (DFG) in the framework of SFB 925 – project ID 170620586 and the excellence cluster `Advanced Imaging of Matter’ - EXC 2056 - project ID 390715994. The project is co-financed by ERDF of the European Union and by `Fonds of
the Hamburg Ministry of Science, Research, Equalities and Districts (BWFGB)’.

E.C. performed the experiments and analyzed data with the help of A.B., E.R. and S.S.
V.P.S. and L.M. developed numerical models. E.C. and V.P.S. performed the numerical simulations. 
E.C., V.P.S. and S.S. wrote the manuscript. 
L.M., C.J.F. and S.S. supervised the project. 
All authors contributed to the discussion and interpretation of our results.

\providecommand{\theHequation}{} 
\providecommand{\theHfigure}{}
\renewcommand{\theHequation}{S\arabic{equation}}
\renewcommand{\theHfigure}{S\arabic{figure}}

\section*{Appendices}
\appendix
\begin{figure}[tb]
    \centering
     \includegraphics[width=\linewidth]{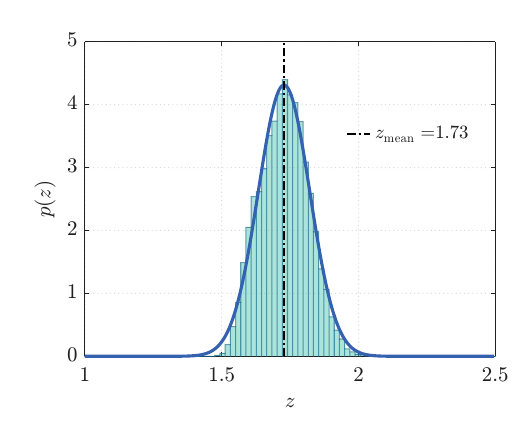}
 \caption{Determination of dynamic critical exponent. Histogram of bootstrapped estimates of the critical exponent $z$ obtained from power‑law fits $f(t) = A \times t^{-2/z}$ to the universal evolution of vortex density as a function of $t/t_c$ across all initial conditions. The blue solid line indicates the Gaussian fit to the observed probability density distribution and the black dash‑dotted line marks the mean value $z=1.73$. }
 \label{fig:bootstrap_fitting}
\end{figure}

\section{Numerical Simulations}
We simulate the many-body dynamics of bilayer 2D Bose gases using classical-field simulations within the  truncated Wigner approximation~\cite{singh_sound_2020,singh_first_2022}. 
For numerical simulations, we discretize real space on a 2D square lattice and represent the continuous Hamiltonian using the discrete Bose-Hubbard Hamiltonian. The system consists of two subsystems (labeled $a=1, 2$) and is described by the Hamiltonian 
\begin{align}
H  = H_1 + H_2 + H_{12}, 
\end{align}
with
\begin{align}\label{eq:Hamil}
H_a &= - J_h \sum_{ \langle i j \rangle } \bigl( \psi_{a, i}^\ast  \psi_{a, j} +  \psi_{a, j}^\ast  \psi_{a, i}  \bigr) + \frac{U}{2} \sum_i n_{a, i}^2 \, \nonumber \\
& \quad  -  \mu  \sum_i   n_{a, i}
\end{align}
and 
\begin{align}
H_{12}= - J \sum_{i} \bigl( \psi_{1, i}^\ast  \psi_{2, i} +  \psi_{2, i}^\ast  \psi_{1, i} \bigr).  
\end{align}
Here, $\langle i j \rangle$ denotes nearest neighbors, $\psi_{a, i}$ and $n_{a, i}= |\psi_{a, i}|^2$ are the complex-valued field and the density at site $i$, respectively.  
$J_h = \hbar^2/(2ml^2) $ is the hopping energy and $U= g/l^2$ is the repulsive interaction energy, 
where $m$ is the mass and $l$ is the discretization length. 
The interaction $g=\tilde{g} \hbar^2/m$ is given in terms of the dimensionless parameter $\tilde{g} =  \sqrt{8 \pi} a_s/\ell_z$, 
where $m$ is the mass, $a_s$ is the $s$-wave scattering length and $\ell_z= \sqrt{\hbar/(m \omega_z)}$ is the harmonic oscillator length in the axial direction. The initial axial trapping frequency is $\omega_z/(2\pi)= 1.6\, \mathrm{kHz}$, resulting in $\tilde{g}=0.098$ for $^{87}$Rb atoms. $H_{12}$ is the coupling Hamiltonian, with $J$ denoting the coupling strength between the two clouds, which is initially set to zero. 
We choose the simulation parameters such as the interaction, phase-space density, and coupling strength according to the experiments.
We consider a lattice system of $152 \times 152$ sites with a discretization length of $l=0.3\, \mu \mathrm{m}$.
We note that for the continuum limit, $l$ is chosen to be smaller than or comparable to the healing length and the thermal de Broglie wavelength $\lambda=h/\sqrt{ 2\pi m k_\mathrm{B} T}$, where $T$ is the temperature, $k_\mathrm{B}$ the Boltzmann constant, and $h$ is the Planck constant~\cite{mora_extension_2003}. 

In  the classical-field approximation, we replace the operators $\hat{\psi}$ by complex numbers $\psi$ both in the Hamiltonian 
and in the equations of motion.
We sample the initial states $\psi_a(\mathbf{r})$ of the decoupled system within a grand-canonical ensemble having temperature $T$ and chemical potential $\mu$  via a classical Metropolis algorithm~\cite{singh_sound_2020,singh_first_2022}.  
We set $T=28\, \mathrm{nK}$ and adjust $\mu$ such that the phase-space density $\mathcal{D} = n \lambda^2$ varies in the range between $7.8$ and $9.3$,  where $n$ is the average 2D density.  
The resulting distribution $\psi_a(\mathbf{r})$ includes fluctuations of the bosonic field beyond mean field.
Finally, each state is propagated using equations of motion
\begin{align}
i \hbar \dot{\psi}_{1, i} &= - J_h \sum_{ \langle i j \rangle }  \psi_{1, i}(j) + U n_{1, i} \psi_{1, i} - J \psi_{2, i}    \\
i \hbar \dot{\psi}_{2, i} &= - J_h \sum_{ \langle i j \rangle }  \psi_{2, i}(j) + U n_{2, i} \psi_{2, i} - J \psi_{1, i}. 
\end{align}
In the time evolution, we ramp up $J$ from $0$ to  $34\, \mathrm{Hz}$ over the duration of $15\, \mathrm{ms}$. 
This also includes a shift of the axial trap frequency from $\omega_z/(2\pi)= 1.6\, \mathrm{kHz}$ to $1.2\, \mathrm{kHz}$,  
resulting in final $\tilde{g} = 0.085$. We examine the resulting dynamics of the system phases $\phi_1(\mathbf{r},t)$ and $\phi_2(\mathbf{r},t)$ of the two clouds, 
which directly gives access to the relative-phase mode via  $\theta(\mathbf{r},t) = \phi_1(\mathbf{r},t) - \phi_2(\mathbf{r},t)$. 
From the phase evolution of the relative mode we compute the two-point correlation function and vortex density, following the same method as in the experiments. Finally, for each value of the  phase-space density, we average the correlation function and the vortex density over the initial ensemble.

\section{Effective low energy description of coupled bilayer 2D Bose gas}
\label{app:effective_low_energy}

We consider a many-body system consisting of a quasi-2D ultracold gas of $^{87}$Rb confined in a double-well potential. The resulting tunnel-coupled 2D system is described by the Hamiltonian 
\begin{equation}
    \begin{aligned}
H = &\int d^2 r\sum_{j=1}^{2}\Big[\frac{\hbar^{2}}{2 m} \nabla \psi_{j}^{\dagger} \cdot \nabla \psi_{j}
+ \frac{g}{2} \psi_{j}^{\dagger} \psi_{j}^{\dagger} \psi_{j} \psi_{j} \\
&+  \left(V-\mu\right) \psi_{j}^{\dagger} \psi_{j}\Big] - \hbar J\int d^2 r\left(\psi_{1}^{\dagger} \psi_{2}+\psi_{2}^{\dagger} \psi_{1}\right),
\end{aligned}
\end{equation}
where $\psi_{j}$ and $ \psi_{j}^{\dagger}$ are bosonic annihilation and creation operators in $j$th layer, respectively, satisfying the commutation relations $[\psi_{j}(\boldsymbol{r}), \psi_{j^{\prime}}^{\dagger}\left(\boldsymbol{r}^{\prime}\right)]=\delta_{j j^{\prime}} \delta^{(2)}\left(\boldsymbol{r}-\boldsymbol{r}^{\prime}\right)$. $g=\tilde{g}\hbar^2/m$ is the quasi-2D interaction strength, $V$ is the trapping potential, $\mu$ is the chemical potential, and  $J$ is the interlayer tunnel coupling strength. $\nabla$ is the 2D gradient in the $\left(x,y\right)$ plane. We then adopt density-phase representation 
\begin{equation}
    \psi_{j}(\boldsymbol{r})=\exp \left[i \theta_{j}(\boldsymbol{r})\right] \sqrt{n_{2 \mathrm{D}}+\delta \rho_{j}(\boldsymbol{r})},
\end{equation}
by assuming a box trapping potential $V = 0$. Phase $\theta_{j}(\boldsymbol{r})$ and density fluctuations $\delta\rho_{j}(\boldsymbol{r})$ obey the commutation relations $[\theta_{j}(\boldsymbol{r}), \delta\rho_{j^{\prime}}\left(\boldsymbol{r}^{\prime}\right)]=-i\,\delta_{j j^{\prime}} \delta^{(2)}\left(\boldsymbol{r}-\boldsymbol{r}^{\prime}\right)$.

For a system with equal atomic populations in each layer, the relative (anti-symmetric) mode in the low-temperature regime is effectively described by a Hamiltonian of the form:
\begin{subequations}
\begin{align}
H_\mathrm{rel} &= H_\mathrm{XY} + H_\mathrm{J}, \\[2ex]
H_\mathrm{XY} &= \int d^2 r \, \frac{\hbar^2}{2 m} \left[
    \frac{\left(\nabla \delta \rho_a\right)^2}{2n_{2 \mathrm{D}}}
    + 2\tilde{g} \delta \rho_a^2
    + \frac{n_{2 \mathrm{D}}}{2}\left(\nabla \theta\right)^2
\right], \\[2ex]
H_\mathrm{J} &= -\int d^2 r \, n_{2 \mathrm{D}} \hbar J \left[
    2 \cos \theta
    - \frac{\delta \rho_a^2}{n_{2 \mathrm{D}}^2} \cos \theta
    + \frac{\delta \rho_s}{n_{2 \mathrm{D}}} \cos \theta
\right].
\end{align}
\label{eq:Hamiltonian}
\end{subequations}
\begin{figure}[b]
    \centering
     \includegraphics[width=\linewidth]{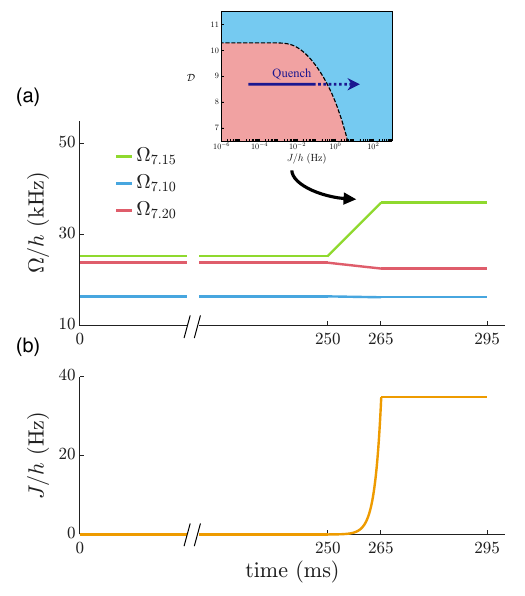}
     \caption{Coupling quench procedure. 
     (a) Time evolution of the RF amplitudes during the coupling quench, expressed as Rabi frequencies $\Omega_{f_i} = g_F \mu_B B_i/2\hbar$ for components $i = 1, 2, 3$, where $B_i$ is the RF magnetic field amplitude at frequency $f_i$, $g_F$ is the Landé $g$-factor, and $\mu_B$ is the Bohr magneton. Atoms are initially confined in a decoupled multiple-RF-dressed two-dimensional (2D) double-well potential and held for an additional \SI{250}{\milli\second} to allow thermal equilibration. The quench is initiated by rapidly increasing the amplitude of $\Omega_{7.15}$ for \SI{15}{\milli\second}, thereby modifying both the barrier height and the separation between the wells. The amplitude of $\Omega_{7.20}$ is dynamically adjusted during the quench to maintain a balanced double-well configuration.
    (b) The inter-well coupling strength exhibits exponential growth during the quench, calculated from two-mode model~\cite{ananikian_grosspitaevskii_2006}. 
}
 \label{fig:ramp}
\end{figure}

Here we apply canonical transformation to express the system in terms of symmetric and antisymmetric degrees of freedom, and expand the Hamiltonian to quadratic order. $\theta(\boldsymbol{r})=\phi_1(\boldsymbol{r})-\phi_2(\boldsymbol{r})$ is the relative (antisymmetric) phase, and the common (symmetric) phase is defined as $\varphi(\boldsymbol{r})=\left(\phi_1(\boldsymbol{r})+\phi_2(\boldsymbol{r})\right)/2$. The density fluctuations in symmetric (subscript s) and antisymmetric (subscript a) are defined as $\delta \rho_{s}(\boldsymbol{r}) = \delta \rho_{1}(\boldsymbol{r}) + \delta \rho_{2}(\boldsymbol{r})$, $\delta \rho_{a}(\boldsymbol{r}) = \left( \delta \rho_{1}(\boldsymbol{r}) - \delta \rho_{2}(\boldsymbol{r})\right)/2$. Notably, the relative phase is coupled to the symmetric degrees of freedom through the tunnel coupling, which can serve as a weak dissipative bath for the relative phase~\cite{dallatorre_universal_2013,vannieuwkerk_lowenergy_2020}. 

In the regime of a quasi-condensate with strongly suppressed density fluctuations, the Hamiltonian for the relative mode reduces to the sine-Gordon form,
\begin{equation}
H_\mathrm{SG} = \int d^2 r \, \frac{\hbar^2}{2 m} \left[
    \frac{n_{2 \mathrm{D}}}{2}\left(\nabla \theta\right)^2+
     2\tilde{g} \delta \rho_a^2 \right] -2n_{2 \mathrm{D}} \hbar J \cos \theta.
\end{equation}

\section{Preparation of non-equilibrium coupled 2D systems}
We begin with an ultra-cold condensate Bose gas of approximately $2 \times 10^4$ ${}^{87}$Rb atoms in the $F = 1$ hyperfine ground state at a temperature of $T = \SI{27}{\nano\kelvin}$, adiabatically loaded into a decoupled double-well quasi-two-dimensional (quasi-2D) potential, as described in detail in Refs.~\cite{harte_ultracold_2018,bentine_2020,sunami_observation_2022}. The potential is engineered using three radio-frequency (RF) fields with frequencies $(f_1, f_2, f_3) = (7.10, 7.15, 7.20)$~MHz, combined with a static quadrupole magnetic field $\bm{B}(\bm{r}) = b(x\bm{e}_x + y\bm{e}_y - 2z\bm{e}_z)$, where the field gradient is $b = \SI{147}{G\,cm^{-1}}$. In addition to the RF-dressed potential, an optical dipole potential is created by a \SI{532}{\nano\meter} ring-shaped laser beam propagating vertically, shaped via direct imaging of a digital micromirror device (DMD) onto the atomic plane. Equal population in the two wells is ensured by maximizing the measured contrast of the matter-wave interference pattern, as described in Ref.~\cite{barker_coherent_2020}.

The atomic clouds are held in the decoupled double-well for 250~ms to allow the system to equilibrate. The barrier height is then rapidly lowered over a duration of \SI{15}{\milli\second} to $E_b/h = \SI{1.94}{\kilo\hertz}$, constituting a non-adiabatic quench for the radial dynamics, which has a characteristic timescale of $2\pi/\omega_r = \SI{86}{\milli\second}$. During the quench, the tunnel coupling strength increases exponentially from $0$ to \SI{34}{\hertz}, as shown in Fig.~\ref{fig:ramp}, while the axial trap frequencies are reduced from $\omega_z/2\pi = \SI{1.6}{\kilo \hertz}$ to $\SI{1.2}{\kilo \hertz}$. The gases are then held in the coupled double-well for a variable duration. In this configuration, a lifetime of \SI{2.2}{\second} is measured, which is sufficiently long to probe the coarsening dynamics reported in the main text.

Following the evolution, the MRF-dressed and optical potentials are turned off to release the clouds into a \SI{16.5}{\milli\second} time-of-flight (TOF), allowing the formation of a matter-wave interference pattern. The optical potential is turned off 3 ms before release to avoid influence during expansion. To image this pattern, atoms are selectively pumped from the $F = 1$ to the $F = 2$ hyperfine state using a repumping light sheet propagating vertically (along the $z$-axis), with a thickness of $L_z = \SI{4}{\micro\metre}$ and a vertical extent much larger than the size of the atomic cloud. The interference pattern is then imaged using light resonant with the $F = 2$ state. We ensure the repumping light intersects the center of the cloud by translating the light sheet along the imaging axis and locating the position that maximizes the total absorption signal.

\begin{figure}[tb]
    \centering
     \includegraphics[width=\linewidth]{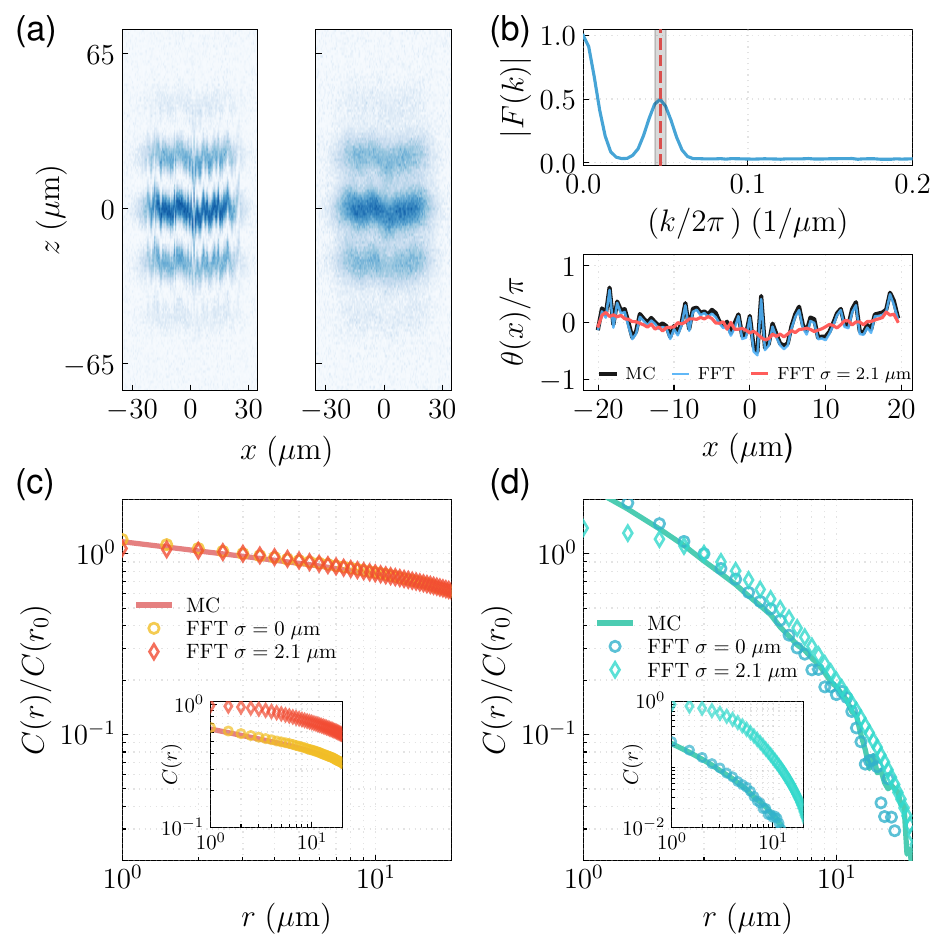}
 \caption{Relative-phase extraction. 
 (a) Simulated interference pattern after TOF with phases generated from Monte Carlo (MC) simulation. The Left displays simulated distribution and the right incorporates the effects of white noise and finite experimental resolution. 
 (b) Top panel shows typical normalized Fourier spectrum of the interference pattern $|F(k)|$ along $z$ axis, the position of peak index $k_0$ is marked by the red dashed line and the shaded region indicates the vicinity of the peak index, $k_0 \pm 1$. 
 In the bottom panel, extracted relative phase $\theta(x)$ is shown alongside the input MC relative phase. 
 (c,d) Phase correlation function obtained from simulated interference pattern (including shot noise of the imaging apparatus), with and without the effect of finite imaging resolution $\sigma$. 
 We rescale the correlation function with its value at fixed distance $C(r_0 = \SI{3}{\micro \meter})$, to visually demonstrate that finite image resolution does not affect the long-distance behavior which is used for the quantitative analysis in this paper. Inset shows the unscaled $C(r)$. 
 We show correlation functions obtained from the simulated interference images at different PSDs, exhibiting power-law and exponential decay for (c) and (d), respectively.}
 \label{fig:FFT}
\end{figure}
\begin{figure*}[tb]
    \centering
     \includegraphics[width=\linewidth]{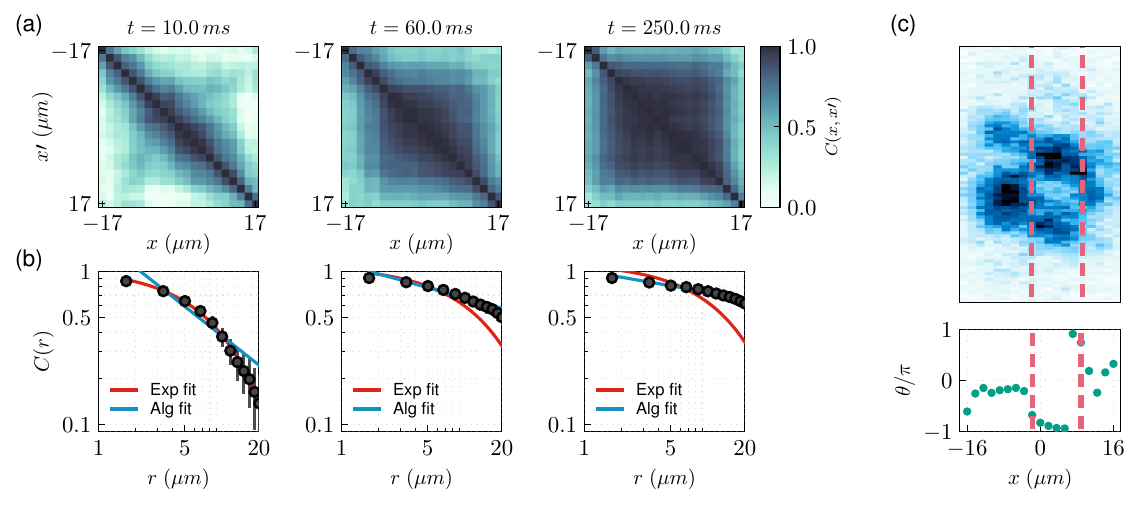}
 \caption{Phase correlation function analysis and vortex detection. 
 (a) Averaged two point correlation $C(x,x^{\prime})$ at different hold times; 
 (b) Spatially averaged correlation function $C(r)$ with $r=x-x^{\prime}$, fitted with exponential and power-law (algebraic) models.  
 (c) Vortices are identified by abrupt phase jumps, indicated by red dashed lines. The bottom panel shows the phase profile extracted from the interference pattern.}
 \label{fig:corr_vortex_analysis}
\end{figure*}

\section{Image analysis for relative phase detection}
The matter-wave interference pattern we observe is of the form~\cite{pethick_bose_2008} 
\begin{equation}\label{eq:fringefit}
    \rho_x(z) = \rho_0 \exp\left(-z^2/2\sigma^2\right) \left[ 1 + c_0 \cos(kz+\theta(x)) \right],
\end{equation}
along $z$ direction at each location $x$. 
The phase $\theta(x)$ encodes a specific realization of the fluctuations of the \textit{in situ} local relative phase. Rather than employing the fitting method to extract the phase~\cite{barker_coherent_2020,hadzibabic_berezinskii_2006}, we use the discrete Fourier transform (DFT), which offers a fast and precise approach for extracting phases from the measured interference patterns~\cite{beregi_thesis} (Fig.~\ref{fig:FFT}). We extract the phase by computing the FFT spectrum averaged over all images and identifying the relevant spatial frequency $k_0$ corresponding to the matter-wave interference pattern. 

\begin{figure}[thb]
    \centering
     \includegraphics[width=\linewidth]{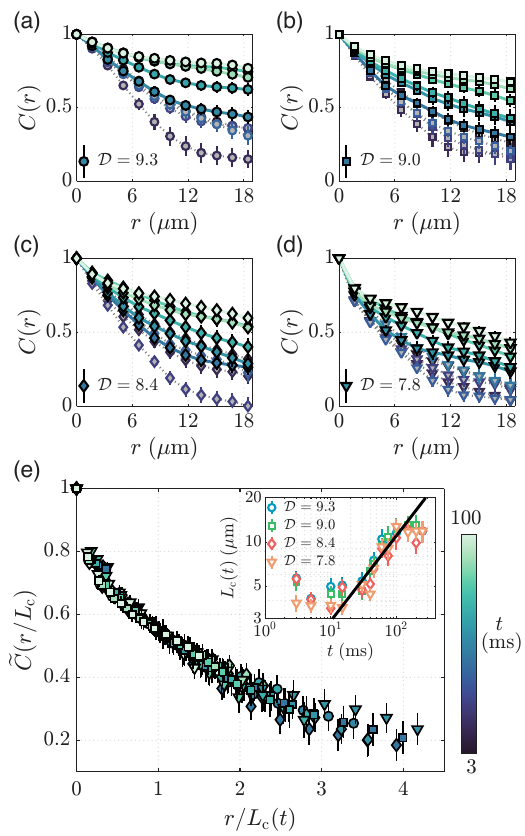}
 \caption{Correlation function dynamics and universal scaling. 
 (a) - (d) 
 Temporal evolution of phase correlation function $C(r)$ for different initial PSDs, $\mathcal{D}$. Data for times $t\lesssim t_c$, are depicted by gray markers connected by  dotted lines. For times  $t \gtrsim t_c$, the evolution is shown by markers with a color gradient, connected by solid lines.
 (e)  Near the critical time and beyond, correlation functions for various initial states collapse onto the universal scaling curve $\widetilde{C}(r/L_c)$. The inset shows the characteristic correlation length $L_c$ for different initial PSDs. The black solid line represents a power-law scaling of $t^{-2/1.72}$.}
 \label{fig:universal_correlation_scaling}
\end{figure}
\begin{figure}[tb]
    \centering
     \includegraphics[width=\linewidth]{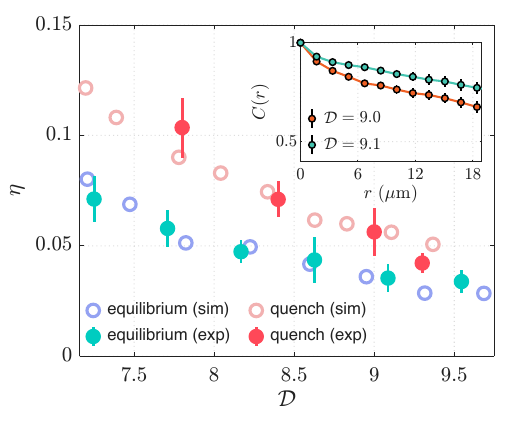}
 \caption{Phase coherence of equilibrium and quenched systems. 
 Extracted algebraic decay exponent \(\eta\) from experiment (filled symbols) and numerical simulation (open symbols) for both the thermal equilibrium and steady states observed after the quench, plotted as a function of PSD. 
 For the quenched systems, as in the main text, the PSD for the equilibrium system before the quench is used for this plot.
 The post‑quench steady state is evaluated at \(t = \SI{250}{\milli\second}\). 
 Inset: Representative correlation functions from experimental data at two nearly identical PSDs for thermal equilibrium and post‑quench steady state.}
 \label{fig:equilibrium}
\end{figure}

For the analysis of the correlation function, phases are obtained by averaging several points in the vicinity of $k_0$, which preserves the phase difference while reducing the impact of noise.  In the experiment, we calculate the averaged phase two-point correlation function $C(x,x^{\prime}) = \Re [ \langle e^{i[\theta(x)-\theta(x^{\prime})]} ]$ from an ensemble of at least 40 images. The averaging is performed over the set of images and different positions $x$ such that $x$ and $x-\overline{x}$ are within the central $\sim\SI{30}{\micro \meter}$ of the density distribution of the cloud. We then calculate $C(r)$ by averaging $C(x,x^{\prime})$ over points with the same spatial separations $r=x-x^{\prime}$, and fit them to both exponential and algebraic models (Fig.~\ref{fig:corr_vortex_analysis}(b)).

Vortices are identified in interference images acquired along the y-direction, following the procedure detailed in~\cite{sunami_observation_2022,sunami_universal_2023}. We detect vortices by searching for abrupt phase jumps within a two-pixel distance ($\SI{3.4}{\micro \meter}$), defined by a phase difference in the range $\pi/3 < \delta \theta < 2\pi/3$ (Fig.~\ref{fig:corr_vortex_analysis}(c)). The vortex density $n_v(x)$ is determined by dividing the probability of finding a vortex in each image column by the effective detection area of a single pixel column $\ell_p L_y =  \SI{6.7}{\micro \metre}^{2}$, where $\ell_p= \SI{1.67}{\micro \metre}$ is the pixel size in the image plane.

\section{Universal scaling for correlation functions}
\label{app:univresal_corr}

In the scaling regime, the correlation function is expected to follow the universal scaling law $C(r, t) \sim A_{\mathcal{D}} F(r/L_c(t))$, where $A_{\mathcal{D}}$ is a non-universal, condition-dependent amplitude and $F(x)$ is the universal scaling function. To demonstrate the universal scaling, we normalize the phase correlation function $\widetilde{C}(r,t) = C(r,t)\times R_{steady}(r,t=\SI{250}{\milli\second})$. The normalization factor, $R_{steady}$, is determined from the ratio of each correlation function, $C(r,t=\SI{250}{\milli\second})$, to that of the reference measurement $\mathcal{D}=7.8$. After normalization, we determine the characteristic correlation length by the condition $\widetilde{C}(L_c(t),t)=0.55$. The result of this analysis is shown in Fig.~\ref{fig:universal_correlation_scaling}(e); for times $t \gtrsim t_c$, all phase correlation functions collapse onto a common curve, which is a clear signature of universal scaling.

\section{Phase coherence of states in equilibrium and steady states after the quench}

We compare the phase coherence between the steady state appearing around 100 ms after the quench and the state in thermal equilibrium. 
To probe the systems in equilibrium, two atomic clouds are adiabatically loaded into the bilayer potential with $J=\SI{34}{\hertz}$ tunnel coupling and held in the trap for a long time to ensure equilibration. 
We then perform matter-wave interferometry (Fig.~\ref{fig:concept}) and obtain the correlation function of the relative phase. 
In Fig.~\ref{fig:equilibrium}, we plot the algebraic exponent $\eta$ for the equilibrium and steady-state samples, where the post‑quench states exhibit systematically larger $\eta$ than the corresponding equilibrium values. This may originate from the sudden quench projecting the initial thermal states onto excited states of the post-quench Hamiltonian, with the associated work injection raising the system’s effective temperature~\cite{mathey_phaselocking_2007,bayocboc_dynamics_2022}. 
We further note from numerical simulations that, compared with the thermal equilibrium state, the common phase relaxes toward a larger $\eta$ on a timescale substantially longer than the $\sim100$ ms relaxation of the relative phase.


\begin{thebibliography}{75}%
\makeatletter
\providecommand \@ifxundefined [1]{%
 \@ifx{#1\undefined}
}%
\providecommand \@ifnum [1]{%
 \ifnum #1\expandafter \@firstoftwo
 \else \expandafter \@secondoftwo
 \fi
}%
\providecommand \@ifx [1]{%
 \ifx #1\expandafter \@firstoftwo
 \else \expandafter \@secondoftwo
 \fi
}%
\providecommand \natexlab [1]{#1}%
\providecommand \enquote  [1]{``#1''}%
\providecommand \bibnamefont  [1]{#1}%
\providecommand \bibfnamefont [1]{#1}%
\providecommand \citenamefont [1]{#1}%
\providecommand \href@noop [0]{\@secondoftwo}%
\providecommand \href [0]{\begingroup \@sanitize@url \@href}%
\providecommand \@href[1]{\@@startlink{#1}\@@href}%
\providecommand \@@href[1]{\endgroup#1\@@endlink}%
\providecommand \@sanitize@url [0]{\catcode `\\12\catcode `\$12\catcode `\&12\catcode `\#12\catcode `\^12\catcode `\_12\catcode `\%12\relax}%
\providecommand \@@startlink[1]{}%
\providecommand \@@endlink[0]{}%
\providecommand \url  [0]{\begingroup\@sanitize@url \@url }%
\providecommand \@url [1]{\endgroup\@href {#1}{\urlprefix }}%
\providecommand \urlprefix  [0]{URL }%
\providecommand \Eprint [0]{\href }%
\providecommand \doibase [0]{https://doi.org/}%
\providecommand \selectlanguage [0]{\@gobble}%
\providecommand \bibinfo  [0]{\@secondoftwo}%
\providecommand \bibfield  [0]{\@secondoftwo}%
\providecommand \translation [1]{[#1]}%
\providecommand \BibitemOpen [0]{}%
\providecommand \bibitemStop [0]{}%
\providecommand \bibitemNoStop [0]{.\EOS\space}%
\providecommand \EOS [0]{\spacefactor3000\relax}%
\providecommand \BibitemShut  [1]{\csname bibitem#1\endcsname}%
\let\auto@bib@innerbib\@empty
\bibitem [{\citenamefont {Micha}\ and\ \citenamefont {Tkachev}(2003)}]{micha_Relativistic_2003}%
  \BibitemOpen
  \bibfield  {author} {\bibinfo {author} {\bibfnamefont {R.}~\bibnamefont {Micha}}\ and\ \bibinfo {author} {\bibfnamefont {I.~I.}\ \bibnamefont {Tkachev}},\ }\bibfield  {title} {\bibinfo {title} {Relativistic {Turbulence}: {A} {Long} {Way} from {Preheating} to {Equilibrium}},\ }\href {https://doi.org/10.1103/PhysRevLett.90.121301} {\bibfield  {journal} {\bibinfo  {journal} {Physical Review Letters}\ }\textbf {\bibinfo {volume} {90}},\ \bibinfo {pages} {121301} (\bibinfo {year} {2003})}\BibitemShut {NoStop}%
\bibitem [{\citenamefont {Kibble}(1976)}]{kibble_topology_1976}%
  \BibitemOpen
  \bibfield  {author} {\bibinfo {author} {\bibfnamefont {T.~W.~B.}\ \bibnamefont {Kibble}},\ }\bibfield  {title} {{\selectlanguage {english}\bibinfo {title} {Topology of cosmic domains and strings}},\ }\href {https://doi.org/10.1088/0305-4470/9/8/029} {\bibfield  {journal} {\bibinfo  {journal} {Journal of Physics A: Mathematical and General}\ }\textbf {\bibinfo {volume} {9}},\ \bibinfo {pages} {1387} (\bibinfo {year} {1976})}\BibitemShut {NoStop}%
\bibitem [{\citenamefont {Zurek}(1985)}]{zurek_cosmological_1985}%
  \BibitemOpen
  \bibfield  {author} {\bibinfo {author} {\bibfnamefont {W.~H.}\ \bibnamefont {Zurek}},\ }\bibfield  {title} {{\selectlanguage {english}\bibinfo {title} {Cosmological experiments in superfluid helium?}},\ }\href {https://doi.org/10.1038/317505a0} {\bibfield  {journal} {\bibinfo  {journal} {Nature}\ }\textbf {\bibinfo {volume} {317}},\ \bibinfo {pages} {505} (\bibinfo {year} {1985})}\BibitemShut {NoStop}%
\bibitem [{\citenamefont {Chuang}\ \emph {et~al.}(1991)\citenamefont {Chuang}, \citenamefont {Durrer}, \citenamefont {Turok},\ and\ \citenamefont {Yurke}}]{chuang_cosmology_1991}%
  \BibitemOpen
  \bibfield  {author} {\bibinfo {author} {\bibfnamefont {I.}~\bibnamefont {Chuang}}, \bibinfo {author} {\bibfnamefont {R.}~\bibnamefont {Durrer}}, \bibinfo {author} {\bibfnamefont {N.}~\bibnamefont {Turok}},\ and\ \bibinfo {author} {\bibfnamefont {B.}~\bibnamefont {Yurke}},\ }\bibfield  {title} {\bibinfo {title} {Cosmology in the {Laboratory}: {Defect} {Dynamics} in {Liquid} {Crystals}},\ }\href {https://doi.org/10.1126/science.251.4999.1336} {\bibfield  {journal} {\bibinfo  {journal} {Science}\ }\textbf {\bibinfo {volume} {251}},\ \bibinfo {pages} {1336} (\bibinfo {year} {1991})}\BibitemShut {NoStop}%
\bibitem [{\citenamefont {Berges}\ \emph {et~al.}(2014)\citenamefont {Berges}, \citenamefont {Boguslavski}, \citenamefont {Schlichting},\ and\ \citenamefont {Venugopalan}}]{berges_turbulent_2014}%
  \BibitemOpen
  \bibfield  {author} {\bibinfo {author} {\bibfnamefont {J.}~\bibnamefont {Berges}}, \bibinfo {author} {\bibfnamefont {K.}~\bibnamefont {Boguslavski}}, \bibinfo {author} {\bibfnamefont {S.}~\bibnamefont {Schlichting}},\ and\ \bibinfo {author} {\bibfnamefont {R.}~\bibnamefont {Venugopalan}},\ }\bibfield  {title} {\bibinfo {title} {Turbulent thermalization process in heavy-ion collisions at ultrarelativistic energies},\ }\href {https://doi.org/10.1103/PhysRevD.89.074011} {\bibfield  {journal} {\bibinfo  {journal} {Physical Review D}\ }\textbf {\bibinfo {volume} {89}},\ \bibinfo {pages} {074011} (\bibinfo {year} {2014})}\BibitemShut {NoStop}%
\bibitem [{\citenamefont {Bray}(1994)}]{bray_theory_1994}%
  \BibitemOpen
  \bibfield  {author} {\bibinfo {author} {\bibfnamefont {A.}~\bibnamefont {Bray}},\ }\bibfield  {title} {\bibinfo {title} {Theory of phase-ordering kinetics},\ }\href {https://doi.org/10.1080/00018739400101505} {\bibfield  {journal} {\bibinfo  {journal} {Advances in Physics}\ }\textbf {\bibinfo {volume} {43}},\ \bibinfo {pages} {357} (\bibinfo {year} {1994})}\BibitemShut {NoStop}%
\bibitem [{\citenamefont {Prüfer}\ \emph {et~al.}(2018)\citenamefont {Prüfer}, \citenamefont {Kunkel}, \citenamefont {Strobel}, \citenamefont {Lannig}, \citenamefont {Linnemann}, \citenamefont {Schmied}, \citenamefont {Berges}, \citenamefont {Gasenzer},\ and\ \citenamefont {Oberthaler}}]{prufer_observation_2018}%
  \BibitemOpen
  \bibfield  {author} {\bibinfo {author} {\bibfnamefont {M.}~\bibnamefont {Prüfer}}, \bibinfo {author} {\bibfnamefont {P.}~\bibnamefont {Kunkel}}, \bibinfo {author} {\bibfnamefont {H.}~\bibnamefont {Strobel}}, \bibinfo {author} {\bibfnamefont {S.}~\bibnamefont {Lannig}}, \bibinfo {author} {\bibfnamefont {D.}~\bibnamefont {Linnemann}}, \bibinfo {author} {\bibfnamefont {C.-M.}\ \bibnamefont {Schmied}}, \bibinfo {author} {\bibfnamefont {J.}~\bibnamefont {Berges}}, \bibinfo {author} {\bibfnamefont {T.}~\bibnamefont {Gasenzer}},\ and\ \bibinfo {author} {\bibfnamefont {M.~K.}\ \bibnamefont {Oberthaler}},\ }\bibfield  {title} {{\selectlanguage {english}\bibinfo {title} {Observation of universal dynamics in a spinor {Bose} gas far from equilibrium}},\ }\href {https://doi.org/10.1038/s41586-018-0659-0} {\bibfield  {journal} {\bibinfo  {journal} {Nature}\ }\textbf {\bibinfo {volume} {563}},\ \bibinfo {pages} {217} (\bibinfo {year} {2018})}\BibitemShut {NoStop}%
\bibitem [{\citenamefont {Mikheev}\ \emph {et~al.}(2023)\citenamefont {Mikheev}, \citenamefont {Siovitz},\ and\ \citenamefont {Gasenzer}}]{mikheev_universal_2023}%
  \BibitemOpen
  \bibfield  {author} {\bibinfo {author} {\bibfnamefont {A.~N.}\ \bibnamefont {Mikheev}}, \bibinfo {author} {\bibfnamefont {I.}~\bibnamefont {Siovitz}},\ and\ \bibinfo {author} {\bibfnamefont {T.}~\bibnamefont {Gasenzer}},\ }\bibfield  {title} {{\selectlanguage {english}\bibinfo {title} {Universal dynamics and non-thermal fixed points in quantum fluids far from equilibrium}},\ }\href {https://doi.org/10.1140/epjs/s11734-023-00974-7} {\bibfield  {journal} {\bibinfo  {journal} {The European Physical Journal Special Topics}\ }\textbf {\bibinfo {volume} {232}},\ \bibinfo {pages} {3393} (\bibinfo {year} {2023})}\BibitemShut {NoStop}%
\bibitem [{\citenamefont {Huh}\ \emph {et~al.}(2024)\citenamefont {Huh}, \citenamefont {Mukherjee}, \citenamefont {Kwon}, \citenamefont {Seo}, \citenamefont {Hur}, \citenamefont {Mistakidis}, \citenamefont {Sadeghpour},\ and\ \citenamefont {Choi}}]{huh_universality_2024}%
  \BibitemOpen
  \bibfield  {author} {\bibinfo {author} {\bibfnamefont {S.}~\bibnamefont {Huh}}, \bibinfo {author} {\bibfnamefont {K.}~\bibnamefont {Mukherjee}}, \bibinfo {author} {\bibfnamefont {K.}~\bibnamefont {Kwon}}, \bibinfo {author} {\bibfnamefont {J.}~\bibnamefont {Seo}}, \bibinfo {author} {\bibfnamefont {J.}~\bibnamefont {Hur}}, \bibinfo {author} {\bibfnamefont {S.~I.}\ \bibnamefont {Mistakidis}}, \bibinfo {author} {\bibfnamefont {H.~R.}\ \bibnamefont {Sadeghpour}},\ and\ \bibinfo {author} {\bibfnamefont {J.-y.}\ \bibnamefont {Choi}},\ }\bibfield  {title} {{\selectlanguage {english}\bibinfo {title} {Universality class of a spinor {Bose}–{Einstein} condensate far from equilibrium}},\ }\href {https://doi.org/10.1038/s41567-023-02339-2} {\bibfield  {journal} {\bibinfo  {journal} {Nature Physics}\ }\textbf {\bibinfo {volume} {20}},\ \bibinfo {pages} {402} (\bibinfo {year} {2024})}\BibitemShut {NoStop}%
\bibitem [{\citenamefont {Lamacraft}(2007)}]{lamacraft_quantum_2007}%
  \BibitemOpen
  \bibfield  {author} {\bibinfo {author} {\bibfnamefont {A.}~\bibnamefont {Lamacraft}},\ }\bibfield  {title} {\bibinfo {title} {Quantum {Quenches} in a {Spinor} {Condensate}},\ }\href {https://doi.org/10.1103/PhysRevLett.98.160404} {\bibfield  {journal} {\bibinfo  {journal} {Physical Review Letters}\ }\textbf {\bibinfo {volume} {98}},\ \bibinfo {pages} {160404} (\bibinfo {year} {2007})}\BibitemShut {NoStop}%
\bibitem [{\citenamefont {Williamson}\ and\ \citenamefont {Blakie}(2016)}]{williamson_universal_2016}%
  \BibitemOpen
  \bibfield  {author} {\bibinfo {author} {\bibfnamefont {L.~A.}\ \bibnamefont {Williamson}}\ and\ \bibinfo {author} {\bibfnamefont {P.}~\bibnamefont {Blakie}},\ }\bibfield  {title} {\bibinfo {title} {Universal {Coarsening} {Dynamics} of a {Quenched} {Ferromagnetic} {Spin}-1 {Condensate}},\ }\href {https://doi.org/10.1103/PhysRevLett.116.025301} {\bibfield  {journal} {\bibinfo  {journal} {Physical Review Letters}\ }\textbf {\bibinfo {volume} {116}},\ \bibinfo {pages} {025301} (\bibinfo {year} {2016})}\BibitemShut {NoStop}%
\bibitem [{\citenamefont {Bourges}\ and\ \citenamefont {Blakie}(2017)}]{bourges_different_2017}%
  \BibitemOpen
  \bibfield  {author} {\bibinfo {author} {\bibfnamefont {A.}~\bibnamefont {Bourges}}\ and\ \bibinfo {author} {\bibfnamefont {P.~B.}\ \bibnamefont {Blakie}},\ }\bibfield  {title} {\bibinfo {title} {Different growth rates for spin and superfluid order in a quenched spinor condensate},\ }\href {https://doi.org/10.1103/PhysRevA.95.023616} {\bibfield  {journal} {\bibinfo  {journal} {Physical Review A}\ }\textbf {\bibinfo {volume} {95}},\ \bibinfo {pages} {023616} (\bibinfo {year} {2017})}\BibitemShut {NoStop}%
\bibitem [{\citenamefont {Mukerjee}\ \emph {et~al.}(2007)\citenamefont {Mukerjee}, \citenamefont {Xu},\ and\ \citenamefont {Moore}}]{mukerjee_dynamical_2007}%
  \BibitemOpen
  \bibfield  {author} {\bibinfo {author} {\bibfnamefont {S.}~\bibnamefont {Mukerjee}}, \bibinfo {author} {\bibfnamefont {C.}~\bibnamefont {Xu}},\ and\ \bibinfo {author} {\bibfnamefont {J.~E.}\ \bibnamefont {Moore}},\ }\bibfield  {title} {\bibinfo {title} {Dynamical models and the phase ordering kinetics of the \$s=1\$ spinor condensate},\ }\href {https://doi.org/10.1103/PhysRevB.76.104519} {\bibfield  {journal} {\bibinfo  {journal} {Physical Review B}\ }\textbf {\bibinfo {volume} {76}},\ \bibinfo {pages} {104519} (\bibinfo {year} {2007})}\BibitemShut {NoStop}%
\bibitem [{\citenamefont {Schmied}\ \emph {et~al.}(2019)\citenamefont {Schmied}, \citenamefont {Gasenzer},\ and\ \citenamefont {Blakie}}]{schmied_violation_2019}%
  \BibitemOpen
  \bibfield  {author} {\bibinfo {author} {\bibfnamefont {C.-M.}\ \bibnamefont {Schmied}}, \bibinfo {author} {\bibfnamefont {T.}~\bibnamefont {Gasenzer}},\ and\ \bibinfo {author} {\bibfnamefont {P.~B.}\ \bibnamefont {Blakie}},\ }\bibfield  {title} {\bibinfo {title} {Violation of single-length-scaling dynamics via spin vortices in an isolated spin-1 {Bose} gas},\ }\href {https://doi.org/10.1103/PhysRevA.100.033603} {\bibfield  {journal} {\bibinfo  {journal} {Physical Review A}\ }\textbf {\bibinfo {volume} {100}},\ \bibinfo {pages} {033603} (\bibinfo {year} {2019})}\BibitemShut {NoStop}%
\bibitem [{\citenamefont {Williamson}\ and\ \citenamefont {Blakie}(2017)}]{williamson_coarsening_2017}%
  \BibitemOpen
  \bibfield  {author} {\bibinfo {author} {\bibfnamefont {L.}~\bibnamefont {Williamson}}\ and\ \bibinfo {author} {\bibfnamefont {P.}~\bibnamefont {Blakie}},\ }\bibfield  {title} {\bibinfo {title} {Coarsening {Dynamics} of an {Isotropic} {Ferromagnetic} {Superfluid}},\ }\href {https://doi.org/10.1103/PhysRevLett.119.255301} {\bibfield  {journal} {\bibinfo  {journal} {Physical Review Letters}\ }\textbf {\bibinfo {volume} {119}},\ \bibinfo {pages} {255301} (\bibinfo {year} {2017})}\BibitemShut {NoStop}%
\bibitem [{\citenamefont {Hofmann}\ \emph {et~al.}(2014)\citenamefont {Hofmann}, \citenamefont {Natu},\ and\ \citenamefont {Das~Sarma}}]{hofmann_coarsening_2014}%
  \BibitemOpen
  \bibfield  {author} {\bibinfo {author} {\bibfnamefont {J.}~\bibnamefont {Hofmann}}, \bibinfo {author} {\bibfnamefont {S.~S.}\ \bibnamefont {Natu}},\ and\ \bibinfo {author} {\bibfnamefont {S.}~\bibnamefont {Das~Sarma}},\ }\bibfield  {title} {\bibinfo {title} {Coarsening {Dynamics} of {Binary} {Bose} {Condensates}},\ }\href {https://doi.org/10.1103/PhysRevLett.113.095702} {\bibfield  {journal} {\bibinfo  {journal} {Physical Review Letters}\ }\textbf {\bibinfo {volume} {113}},\ \bibinfo {pages} {095702} (\bibinfo {year} {2014})}\BibitemShut {NoStop}%
\bibitem [{\citenamefont {Fujimoto}\ \emph {et~al.}(2020)\citenamefont {Fujimoto}, \citenamefont {Haneda}, \citenamefont {Kudo},\ and\ \citenamefont {Kawaguchi}}]{fujimoto_scaleinvariant_2020}%
  \BibitemOpen
  \bibfield  {author} {\bibinfo {author} {\bibfnamefont {K.}~\bibnamefont {Fujimoto}}, \bibinfo {author} {\bibfnamefont {K.}~\bibnamefont {Haneda}}, \bibinfo {author} {\bibfnamefont {K.}~\bibnamefont {Kudo}},\ and\ \bibinfo {author} {\bibfnamefont {Y.}~\bibnamefont {Kawaguchi}},\ }\bibfield  {title} {\bibinfo {title} {Scale-invariant relaxation dynamics in two-component {Bose}-{Einstein} condensates with large particle-number imbalance},\ }\href {https://doi.org/10.1103/PhysRevA.101.023608} {\bibfield  {journal} {\bibinfo  {journal} {Physical Review A}\ }\textbf {\bibinfo {volume} {101}},\ \bibinfo {pages} {023608} (\bibinfo {year} {2020})}\BibitemShut {NoStop}%
\bibitem [{\citenamefont {Singh}\ \emph {et~al.}(2023)\citenamefont {Singh}, \citenamefont {Amico},\ and\ \citenamefont {Mathey}}]{singh_thermal_2023}%
  \BibitemOpen
  \bibfield  {author} {\bibinfo {author} {\bibfnamefont {V.~P.}\ \bibnamefont {Singh}}, \bibinfo {author} {\bibfnamefont {L.}~\bibnamefont {Amico}},\ and\ \bibinfo {author} {\bibfnamefont {L.}~\bibnamefont {Mathey}},\ }\bibfield  {title} {\bibinfo {title} {Thermal suppression of demixing dynamics in a binary condensate},\ }\href {https://doi.org/10.1103/PhysRevResearch.5.043042} {\bibfield  {journal} {\bibinfo  {journal} {Physical Review Research}\ }\textbf {\bibinfo {volume} {5}},\ \bibinfo {pages} {043042} (\bibinfo {year} {2023})}\BibitemShut {NoStop}%
\bibitem [{\citenamefont {{Rajat}}\ \emph {et~al.}(2025)\citenamefont {{Rajat}}, \citenamefont {Banger},\ and\ \citenamefont {Gautam}}]{rajat_collective_2025}%
  \BibitemOpen
  \bibfield  {author} {\bibinfo {author} {\bibnamefont {{Rajat}}}, \bibinfo {author} {\bibfnamefont {P.}~\bibnamefont {Banger}},\ and\ \bibinfo {author} {\bibfnamefont {S.}~\bibnamefont {Gautam}},\ }\bibfield  {title} {\bibinfo {title} {Collective excitations and universal coarsening dynamics of a spin-orbit-coupled spin-1 {Bose}-{Einstein} condensate},\ }\href {https://link.aps.org/doi/10.1103/PhysRevA.111.033316} {\bibfield  {journal} {\bibinfo  {journal} {Physical Review A}\ }\textbf {\bibinfo {volume} {111}},\ \bibinfo {pages} {033316} (\bibinfo {year} {2025})}\BibitemShut {NoStop}%
\bibitem [{\citenamefont {Kulczykowski}\ and\ \citenamefont {Matuszewski}(2017)}]{kulczykowski_phase_2017}%
  \BibitemOpen
  \bibfield  {author} {\bibinfo {author} {\bibfnamefont {M.}~\bibnamefont {Kulczykowski}}\ and\ \bibinfo {author} {\bibfnamefont {M.}~\bibnamefont {Matuszewski}},\ }\bibfield  {title} {\bibinfo {title} {Phase ordering kinetics of a nonequilibrium exciton-polariton condensate},\ }\href {https://doi.org/10.1103/PhysRevB.95.075306} {\bibfield  {journal} {\bibinfo  {journal} {Physical Review B}\ }\textbf {\bibinfo {volume} {95}},\ \bibinfo {pages} {075306} (\bibinfo {year} {2017})}\BibitemShut {NoStop}%
\bibitem [{\citenamefont {Comaron}\ \emph {et~al.}(2018)\citenamefont {Comaron}, \citenamefont {Dagvadorj}, \citenamefont {Zamora}, \citenamefont {Carusotto}, \citenamefont {Proukakis},\ and\ \citenamefont {Szymańska}}]{comaron_dynamical_2018}%
  \BibitemOpen
  \bibfield  {author} {\bibinfo {author} {\bibfnamefont {P.}~\bibnamefont {Comaron}}, \bibinfo {author} {\bibfnamefont {G.}~\bibnamefont {Dagvadorj}}, \bibinfo {author} {\bibfnamefont {A.}~\bibnamefont {Zamora}}, \bibinfo {author} {\bibfnamefont {I.}~\bibnamefont {Carusotto}}, \bibinfo {author} {\bibfnamefont {N.}~\bibnamefont {Proukakis}},\ and\ \bibinfo {author} {\bibfnamefont {M.}~\bibnamefont {Szymańska}},\ }\bibfield  {title} {\bibinfo {title} {Dynamical {Critical} {Exponents} in {Driven}-{Dissipative} {Quantum} {Systems}},\ }\href {https://doi.org/10.1103/PhysRevLett.121.095302} {\bibfield  {journal} {\bibinfo  {journal} {Physical Review Letters}\ }\textbf {\bibinfo {volume} {121}},\ \bibinfo {pages} {095302} (\bibinfo {year} {2018})}\BibitemShut {NoStop}%
\bibitem [{\citenamefont {Gladilin}\ and\ \citenamefont {Wouters}(2019)}]{gladilin_multivortex_2019}%
  \BibitemOpen
  \bibfield  {author} {\bibinfo {author} {\bibfnamefont {V.~N.}\ \bibnamefont {Gladilin}}\ and\ \bibinfo {author} {\bibfnamefont {M.}~\bibnamefont {Wouters}},\ }\bibfield  {title} {{\selectlanguage {english}\bibinfo {title} {Multivortex states and dynamics in nonequilibrium polariton condensates}},\ }\href {https://doi.org/10.1088/1751-8121/ab3abc} {\bibfield  {journal} {\bibinfo  {journal} {Journal of Physics A: Mathematical and Theoretical}\ }\textbf {\bibinfo {volume} {52}},\ \bibinfo {pages} {395303} (\bibinfo {year} {2019})}\BibitemShut {NoStop}%
\bibitem [{\citenamefont {Manovitz}\ \emph {et~al.}(2025)\citenamefont {Manovitz}, \citenamefont {Li}, \citenamefont {Ebadi}, \citenamefont {Samajdar}, \citenamefont {Geim}, \citenamefont {Evered}, \citenamefont {Bluvstein}, \citenamefont {Zhou}, \citenamefont {Koyluoglu}, \citenamefont {Feldmeier}, \citenamefont {Dolgirev}, \citenamefont {Maskara}, \citenamefont {Kalinowski}, \citenamefont {Sachdev}, \citenamefont {Huse}, \citenamefont {Greiner}, \citenamefont {Vuletić},\ and\ \citenamefont {Lukin}}]{manovitz_quantum_2025}%
  \BibitemOpen
  \bibfield  {author} {\bibinfo {author} {\bibfnamefont {T.}~\bibnamefont {Manovitz}}, \bibinfo {author} {\bibfnamefont {S.~H.}\ \bibnamefont {Li}}, \bibinfo {author} {\bibfnamefont {S.}~\bibnamefont {Ebadi}}, \bibinfo {author} {\bibfnamefont {R.}~\bibnamefont {Samajdar}}, \bibinfo {author} {\bibfnamefont {A.~A.}\ \bibnamefont {Geim}}, \bibinfo {author} {\bibfnamefont {S.~J.}\ \bibnamefont {Evered}}, \bibinfo {author} {\bibfnamefont {D.}~\bibnamefont {Bluvstein}}, \bibinfo {author} {\bibfnamefont {H.}~\bibnamefont {Zhou}}, \bibinfo {author} {\bibfnamefont {N.~U.}\ \bibnamefont {Koyluoglu}}, \bibinfo {author} {\bibfnamefont {J.}~\bibnamefont {Feldmeier}}, \bibinfo {author} {\bibfnamefont {P.~E.}\ \bibnamefont {Dolgirev}}, \bibinfo {author} {\bibfnamefont {N.}~\bibnamefont {Maskara}}, \bibinfo {author} {\bibfnamefont {M.}~\bibnamefont {Kalinowski}}, \bibinfo {author} {\bibfnamefont {S.}~\bibnamefont {Sachdev}}, \bibinfo {author} {\bibfnamefont {D.~A.}\ \bibnamefont {Huse}}, \bibinfo {author} {\bibfnamefont
  {M.}~\bibnamefont {Greiner}}, \bibinfo {author} {\bibfnamefont {V.}~\bibnamefont {Vuletić}},\ and\ \bibinfo {author} {\bibfnamefont {M.~D.}\ \bibnamefont {Lukin}},\ }\bibfield  {title} {{\selectlanguage {english}\bibinfo {title} {Quantum coarsening and collective dynamics on a programmable simulator}},\ }\href {https://doi.org/10.1038/s41586-024-08353-5} {\bibfield  {journal} {\bibinfo  {journal} {Nature}\ }\textbf {\bibinfo {volume} {638}},\ \bibinfo {pages} {86} (\bibinfo {year} {2025})}\BibitemShut {NoStop}%
\bibitem [{\citenamefont {Gazo}\ \emph {et~al.}(2025)\citenamefont {Gazo}, \citenamefont {Karailiev}, \citenamefont {Satoor}, \citenamefont {Eigen}, \citenamefont {Gałka},\ and\ \citenamefont {Hadzibabic}}]{gazo_universal_2025}%
  \BibitemOpen
  \bibfield  {author} {\bibinfo {author} {\bibfnamefont {M.}~\bibnamefont {Gazo}}, \bibinfo {author} {\bibfnamefont {A.}~\bibnamefont {Karailiev}}, \bibinfo {author} {\bibfnamefont {T.}~\bibnamefont {Satoor}}, \bibinfo {author} {\bibfnamefont {C.}~\bibnamefont {Eigen}}, \bibinfo {author} {\bibfnamefont {M.}~\bibnamefont {Gałka}},\ and\ \bibinfo {author} {\bibfnamefont {Z.}~\bibnamefont {Hadzibabic}},\ }\bibfield  {title} {\bibinfo {title} {Universal coarsening in a homogeneous two-dimensional {Bose} gas},\ }\href {https://doi.org/10.1126/science.ado3487} {\bibfield  {journal} {\bibinfo  {journal} {Science}\ }\textbf {\bibinfo {volume} {389}},\ \bibinfo {pages} {802} (\bibinfo {year} {2025})}\BibitemShut {NoStop}%
\bibitem [{\citenamefont {Karl}\ and\ \citenamefont {Gasenzer}(2017)}]{karl_strongly_2017}%
  \BibitemOpen
  \bibfield  {author} {\bibinfo {author} {\bibfnamefont {M.}~\bibnamefont {Karl}}\ and\ \bibinfo {author} {\bibfnamefont {T.}~\bibnamefont {Gasenzer}},\ }\bibfield  {title} {{\selectlanguage {english}\bibinfo {title} {Strongly anomalous non-thermal fixed point in a quenched two-dimensional {Bose} gas}},\ }\href {https://doi.org/10.1088/1367-2630/aa7eeb} {\bibfield  {journal} {\bibinfo  {journal} {New Journal of Physics}\ }\textbf {\bibinfo {volume} {19}},\ \bibinfo {pages} {093014} (\bibinfo {year} {2017})}\BibitemShut {NoStop}%
\bibitem [{\citenamefont {Comaron}\ \emph {et~al.}(2019)\citenamefont {Comaron}, \citenamefont {Larcher}, \citenamefont {Dalfovo},\ and\ \citenamefont {Proukakis}}]{comaron_quench_2019}%
  \BibitemOpen
  \bibfield  {author} {\bibinfo {author} {\bibfnamefont {P.}~\bibnamefont {Comaron}}, \bibinfo {author} {\bibfnamefont {F.}~\bibnamefont {Larcher}}, \bibinfo {author} {\bibfnamefont {F.}~\bibnamefont {Dalfovo}},\ and\ \bibinfo {author} {\bibfnamefont {N.~P.}\ \bibnamefont {Proukakis}},\ }\bibfield  {title} {\bibinfo {title} {Quench dynamics of an ultracold two-dimensional {Bose} gas},\ }\href {https://doi.org/10.1103/PhysRevA.100.033618} {\bibfield  {journal} {\bibinfo  {journal} {Physical Review A}\ }\textbf {\bibinfo {volume} {100}},\ \bibinfo {pages} {033618} (\bibinfo {year} {2019})}\BibitemShut {NoStop}%
\bibitem [{\citenamefont {Groszek}\ \emph {et~al.}(2021)\citenamefont {Groszek}, \citenamefont {Comaron}, \citenamefont {Proukakis},\ and\ \citenamefont {Billam}}]{groszek_crossover_2021}%
  \BibitemOpen
  \bibfield  {author} {\bibinfo {author} {\bibfnamefont {A.~J.}\ \bibnamefont {Groszek}}, \bibinfo {author} {\bibfnamefont {P.}~\bibnamefont {Comaron}}, \bibinfo {author} {\bibfnamefont {N.~P.}\ \bibnamefont {Proukakis}},\ and\ \bibinfo {author} {\bibfnamefont {T.~P.}\ \bibnamefont {Billam}},\ }\bibfield  {title} {\bibinfo {title} {Crossover in the dynamical critical exponent of a quenched two-dimensional {Bose} gas},\ }\href {https://doi.org/10.1103/PhysRevResearch.3.013212} {\bibfield  {journal} {\bibinfo  {journal} {Physical Review Research}\ }\textbf {\bibinfo {volume} {3}},\ \bibinfo {pages} {013212} (\bibinfo {year} {2021})}\BibitemShut {NoStop}%
\bibitem [{\citenamefont {Mathey}\ \emph {et~al.}(2007)\citenamefont {Mathey}, \citenamefont {Polkovnikov},\ and\ \citenamefont {Castro~Neto}}]{mathey_phaselocking_2007}%
  \BibitemOpen
  \bibfield  {author} {\bibinfo {author} {\bibfnamefont {L.}~\bibnamefont {Mathey}}, \bibinfo {author} {\bibfnamefont {A.}~\bibnamefont {Polkovnikov}},\ and\ \bibinfo {author} {\bibfnamefont {A.~H.}\ \bibnamefont {Castro~Neto}},\ }\bibfield  {title} {{\selectlanguage {english}\bibinfo {title} {Phase-locking transition of coupled low-dimensional superfluids}},\ }\href {https://doi.org/10.1209/0295-5075/81/10008} {\bibfield  {journal} {\bibinfo  {journal} {Europhysics Letters}\ }\textbf {\bibinfo {volume} {81}},\ \bibinfo {pages} {10008} (\bibinfo {year} {2007})}\BibitemShut {NoStop}%
\bibitem [{\citenamefont {Rydow}\ \emph {et~al.}(2025)\citenamefont {Rydow}, \citenamefont {Singh}, \citenamefont {Beregi}, \citenamefont {Chang}, \citenamefont {Mathey}, \citenamefont {Foot},\ and\ \citenamefont {Sunami}}]{rydow_observation_2025}%
  \BibitemOpen
  \bibfield  {author} {\bibinfo {author} {\bibfnamefont {E.}~\bibnamefont {Rydow}}, \bibinfo {author} {\bibfnamefont {V.~P.}\ \bibnamefont {Singh}}, \bibinfo {author} {\bibfnamefont {A.}~\bibnamefont {Beregi}}, \bibinfo {author} {\bibfnamefont {E.}~\bibnamefont {Chang}}, \bibinfo {author} {\bibfnamefont {L.}~\bibnamefont {Mathey}}, \bibinfo {author} {\bibfnamefont {C.~J.}\ \bibnamefont {Foot}},\ and\ \bibinfo {author} {\bibfnamefont {S.}~\bibnamefont {Sunami}},\ }\bibfield  {title} {\bibinfo {title} {Observation of a bilayer superfluid with interlayer coherence},\ }\href {https://doi.org/10.1038/s41467-025-62277-w} {\bibfield  {journal} {\bibinfo  {journal} {Nature Communications}\ }\textbf {\bibinfo {volume} {16}},\ \bibinfo {pages} {7201} (\bibinfo {year} {2025})}\BibitemShut {NoStop}%
\bibitem [{\citenamefont {Zhang}\ and\ \citenamefont {Fertig}(2005)}]{zhang_vortices_2005}%
  \BibitemOpen
  \bibfield  {author} {\bibinfo {author} {\bibfnamefont {W.}~\bibnamefont {Zhang}}\ and\ \bibinfo {author} {\bibfnamefont {H.~A.}\ \bibnamefont {Fertig}},\ }\bibfield  {title} {\bibinfo {title} {Vortices and dissipation in a bilayer thin film superconductor},\ }\href {https://doi.org/10.1103/PhysRevB.71.224514} {\bibfield  {journal} {\bibinfo  {journal} {Physical Review B}\ }\textbf {\bibinfo {volume} {71}},\ \bibinfo {pages} {224514} (\bibinfo {year} {2005})}\BibitemShut {NoStop}%
\bibitem [{\citenamefont {Leggett}(2006)}]{leggett_what_2006}%
  \BibitemOpen
  \bibfield  {author} {\bibinfo {author} {\bibfnamefont {A.~J.}\ \bibnamefont {Leggett}},\ }\bibfield  {title} {{\selectlanguage {english}\bibinfo {title} {What {DO} we know about high {Tc}?}},\ }\href {https://doi.org/10.1038/nphys254} {\bibfield  {journal} {\bibinfo  {journal} {Nature Physics}\ }\textbf {\bibinfo {volume} {2}},\ \bibinfo {pages} {134} (\bibinfo {year} {2006})}\BibitemShut {NoStop}%
\bibitem [{\citenamefont {Benfatto}\ \emph {et~al.}(2007)\citenamefont {Benfatto}, \citenamefont {Castellani},\ and\ \citenamefont {Giamarchi}}]{benfatto_kosterlitzthouless_2007}%
  \BibitemOpen
  \bibfield  {author} {\bibinfo {author} {\bibfnamefont {L.}~\bibnamefont {Benfatto}}, \bibinfo {author} {\bibfnamefont {C.}~\bibnamefont {Castellani}},\ and\ \bibinfo {author} {\bibfnamefont {T.}~\bibnamefont {Giamarchi}},\ }\bibfield  {title} {\bibinfo {title} {Kosterlitz-{Thouless} {Behavior} in {Layered} {Superconductors}: {The} {Role} of the {Vortex} {Core} {Energy}},\ }\href {https://doi.org/10.1103/PhysRevLett.98.117008} {\bibfield  {journal} {\bibinfo  {journal} {Physical Review Letters}\ }\textbf {\bibinfo {volume} {98}},\ \bibinfo {pages} {117008} (\bibinfo {year} {2007})}\BibitemShut {NoStop}%
\bibitem [{\citenamefont {Okamoto}\ \emph {et~al.}(2017)\citenamefont {Okamoto}, \citenamefont {Hu}, \citenamefont {Cavalleri},\ and\ \citenamefont {Mathey}}]{okamoto2017}%
  \BibitemOpen
  \bibfield  {author} {\bibinfo {author} {\bibfnamefont {J.-i.}\ \bibnamefont {Okamoto}}, \bibinfo {author} {\bibfnamefont {W.}~\bibnamefont {Hu}}, \bibinfo {author} {\bibfnamefont {A.}~\bibnamefont {Cavalleri}},\ and\ \bibinfo {author} {\bibfnamefont {L.}~\bibnamefont {Mathey}},\ }\bibfield  {title} {\bibinfo {title} {Transiently enhanced interlayer tunneling in optically driven high-${T}_{c}$ superconductors},\ }\href {https://doi.org/10.1103/PhysRevB.96.144505} {\bibfield  {journal} {\bibinfo  {journal} {Phys. Rev. B}\ }\textbf {\bibinfo {volume} {96}},\ \bibinfo {pages} {144505} (\bibinfo {year} {2017})}\BibitemShut {NoStop}%
\bibitem [{\citenamefont {Homann}\ \emph {et~al.}(2024)\citenamefont {Homann}, \citenamefont {Michael}, \citenamefont {Cosme},\ and\ \citenamefont {Mathey}}]{homann_dissipationless_2024}%
  \BibitemOpen
  \bibfield  {author} {\bibinfo {author} {\bibfnamefont {G.}~\bibnamefont {Homann}}, \bibinfo {author} {\bibfnamefont {M.~H.}\ \bibnamefont {Michael}}, \bibinfo {author} {\bibfnamefont {J.~G.}\ \bibnamefont {Cosme}},\ and\ \bibinfo {author} {\bibfnamefont {L.}~\bibnamefont {Mathey}},\ }\bibfield  {title} {\bibinfo {title} {Dissipationless {Counterflow} {Currents} above \$\{{T}\}\_\{c\}\$ in {Bilayer} {Superconductors}},\ }\href {https://doi.org/10.1103/PhysRevLett.132.096002} {\bibfield  {journal} {\bibinfo  {journal} {Physical Review Letters}\ }\textbf {\bibinfo {volume} {132}},\ \bibinfo {pages} {096002} (\bibinfo {year} {2024})}\BibitemShut {NoStop}%
\bibitem [{\citenamefont {Eto}\ and\ \citenamefont {Nitta}(2018)}]{Eto2018}%
  \BibitemOpen
  \bibfield  {author} {\bibinfo {author} {\bibfnamefont {M.}~\bibnamefont {Eto}}\ and\ \bibinfo {author} {\bibfnamefont {M.}~\bibnamefont {Nitta}},\ }\bibfield  {title} {\bibinfo {title} {Confinement of half-quantized vortices in coherently coupled bose-einstein condensates: Simulating quark confinement in a qcd-like theory},\ }\href {https://doi.org/10.1103/PhysRevA.97.023613} {\bibfield  {journal} {\bibinfo  {journal} {Phys. Rev. A}\ }\textbf {\bibinfo {volume} {97}},\ \bibinfo {pages} {023613} (\bibinfo {year} {2018})}\BibitemShut {NoStop}%
\bibitem [{\citenamefont {Kobayashi}\ \emph {et~al.}(2019)\citenamefont {Kobayashi}, \citenamefont {Eto},\ and\ \citenamefont {Nitta}}]{Kobayashi2019}%
  \BibitemOpen
  \bibfield  {author} {\bibinfo {author} {\bibfnamefont {M.}~\bibnamefont {Kobayashi}}, \bibinfo {author} {\bibfnamefont {M.}~\bibnamefont {Eto}},\ and\ \bibinfo {author} {\bibfnamefont {M.}~\bibnamefont {Nitta}},\ }\bibfield  {title} {\bibinfo {title} {Berezinskii-kosterlitz-thouless transition of two-component bose mixtures with intercomponent josephson coupling},\ }\href {https://doi.org/10.1103/PhysRevLett.123.075303} {\bibfield  {journal} {\bibinfo  {journal} {Phys. Rev. Lett.}\ }\textbf {\bibinfo {volume} {123}},\ \bibinfo {pages} {075303} (\bibinfo {year} {2019})}\BibitemShut {NoStop}%
\bibitem [{\citenamefont {Tylutki}\ \emph {et~al.}(2016)\citenamefont {Tylutki}, \citenamefont {Pitaevskii}, \citenamefont {Recati},\ and\ \citenamefont {Stringari}}]{Tylutki2016}%
  \BibitemOpen
  \bibfield  {author} {\bibinfo {author} {\bibfnamefont {M.}~\bibnamefont {Tylutki}}, \bibinfo {author} {\bibfnamefont {L.~P.}\ \bibnamefont {Pitaevskii}}, \bibinfo {author} {\bibfnamefont {A.}~\bibnamefont {Recati}},\ and\ \bibinfo {author} {\bibfnamefont {S.}~\bibnamefont {Stringari}},\ }\bibfield  {title} {\bibinfo {title} {Confinement and precession of vortex pairs in coherently coupled bose-einstein condensates},\ }\href {https://doi.org/10.1103/PhysRevA.93.043623} {\bibfield  {journal} {\bibinfo  {journal} {Phys. Rev. A}\ }\textbf {\bibinfo {volume} {93}},\ \bibinfo {pages} {043623} (\bibinfo {year} {2016})}\BibitemShut {NoStop}%
\bibitem [{\citenamefont {Cazalilla}\ \emph {et~al.}(2007)\citenamefont {Cazalilla}, \citenamefont {Iucci},\ and\ \citenamefont {Giamarchi}}]{cazalilla_competition_2007}%
  \BibitemOpen
  \bibfield  {author} {\bibinfo {author} {\bibfnamefont {M.~A.}\ \bibnamefont {Cazalilla}}, \bibinfo {author} {\bibfnamefont {A.}~\bibnamefont {Iucci}},\ and\ \bibinfo {author} {\bibfnamefont {T.}~\bibnamefont {Giamarchi}},\ }\bibfield  {title} {\bibinfo {title} {Competition between vortex unbinding and tunneling in an optical lattice},\ }\href {https://doi.org/10.1103/PhysRevA.75.051603} {\bibfield  {journal} {\bibinfo  {journal} {Physical Review A}\ }\textbf {\bibinfo {volume} {75}},\ \bibinfo {pages} {051603} (\bibinfo {year} {2007})}\BibitemShut {NoStop}%
\bibitem [{\citenamefont {Bighin}\ \emph {et~al.}(2019)\citenamefont {Bighin}, \citenamefont {Defenu}, \citenamefont {Nándori}, \citenamefont {Salasnich},\ and\ \citenamefont {Trombettoni}}]{bighin_berezinskiikosterlitzthouless_2019}%
  \BibitemOpen
  \bibfield  {author} {\bibinfo {author} {\bibfnamefont {G.}~\bibnamefont {Bighin}}, \bibinfo {author} {\bibfnamefont {N.}~\bibnamefont {Defenu}}, \bibinfo {author} {\bibfnamefont {I.}~\bibnamefont {Nándori}}, \bibinfo {author} {\bibfnamefont {L.}~\bibnamefont {Salasnich}},\ and\ \bibinfo {author} {\bibfnamefont {A.}~\bibnamefont {Trombettoni}},\ }\bibfield  {title} {\bibinfo {title} {Berezinskii-{Kosterlitz}-{Thouless} {Paired} {Phase} in {Coupled} \${XY}\$ {Models}},\ }\href {https://doi.org/10.1103/PhysRevLett.123.100601} {\bibfield  {journal} {\bibinfo  {journal} {Physical Review Letters}\ }\textbf {\bibinfo {volume} {123}},\ \bibinfo {pages} {100601} (\bibinfo {year} {2019})}\BibitemShut {NoStop}%
\bibitem [{\citenamefont {Xiao}\ \emph {et~al.}(2025)\citenamefont {Xiao}, \citenamefont {Deng},\ and\ \citenamefont {Dong}}]{xiao_fate_2025}%
  \BibitemOpen
  \bibfield  {author} {\bibinfo {author} {\bibfnamefont {T.}~\bibnamefont {Xiao}}, \bibinfo {author} {\bibfnamefont {Y.}~\bibnamefont {Deng}},\ and\ \bibinfo {author} {\bibfnamefont {X.-Y.}\ \bibnamefont {Dong}},\ }\href {https://doi.org/10.48550/arXiv.2504.01461} {\bibinfo {title} {Fate of {Berezinskii}-{Kosterlitz}-{Thouless} {Paired} {Phase} in {Coupled} \${XY}\$ {Models}}} (\bibinfo {year} {2025}),\ \bibinfo {note} {arXiv:2504.01461 [cond-mat]}\BibitemShut {NoStop}%
\bibitem [{\citenamefont {Dalla~Torre}\ \emph {et~al.}(2013)\citenamefont {Dalla~Torre}, \citenamefont {Demler},\ and\ \citenamefont {Polkovnikov}}]{dallatorre_universal_2013}%
  \BibitemOpen
  \bibfield  {author} {\bibinfo {author} {\bibfnamefont {E.~G.}\ \bibnamefont {Dalla~Torre}}, \bibinfo {author} {\bibfnamefont {E.}~\bibnamefont {Demler}},\ and\ \bibinfo {author} {\bibfnamefont {A.}~\bibnamefont {Polkovnikov}},\ }\bibfield  {title} {\bibinfo {title} {Universal {Rephasing} {Dynamics} after a {Quantum} {Quench} via {Sudden} {Coupling} of {Two} {Initially} {Independent} {Condensates}},\ }\href {https://doi.org/10.1103/PhysRevLett.110.090404} {\bibfield  {journal} {\bibinfo  {journal} {Physical Review Letters}\ }\textbf {\bibinfo {volume} {110}},\ \bibinfo {pages} {090404} (\bibinfo {year} {2013})}\BibitemShut {NoStop}%
\bibitem [{\citenamefont {Schweigler}\ \emph {et~al.}(2017)\citenamefont {Schweigler}, \citenamefont {Kasper}, \citenamefont {Erne}, \citenamefont {Mazets}, \citenamefont {Rauer}, \citenamefont {Cataldini}, \citenamefont {Langen}, \citenamefont {Gasenzer}, \citenamefont {Berges},\ and\ \citenamefont {Schmiedmayer}}]{schweigler_experimental_2017}%
  \BibitemOpen
  \bibfield  {author} {\bibinfo {author} {\bibfnamefont {T.}~\bibnamefont {Schweigler}}, \bibinfo {author} {\bibfnamefont {V.}~\bibnamefont {Kasper}}, \bibinfo {author} {\bibfnamefont {S.}~\bibnamefont {Erne}}, \bibinfo {author} {\bibfnamefont {I.}~\bibnamefont {Mazets}}, \bibinfo {author} {\bibfnamefont {B.}~\bibnamefont {Rauer}}, \bibinfo {author} {\bibfnamefont {F.}~\bibnamefont {Cataldini}}, \bibinfo {author} {\bibfnamefont {T.}~\bibnamefont {Langen}}, \bibinfo {author} {\bibfnamefont {T.}~\bibnamefont {Gasenzer}}, \bibinfo {author} {\bibfnamefont {J.}~\bibnamefont {Berges}},\ and\ \bibinfo {author} {\bibfnamefont {J.}~\bibnamefont {Schmiedmayer}},\ }\bibfield  {title} {{\selectlanguage {english}\bibinfo {title} {Experimental characterization of a quantum many-body system via higher-order correlations}},\ }\href {https://doi.org/10.1038/nature22310} {\bibfield  {journal} {\bibinfo  {journal} {Nature}\ }\textbf {\bibinfo {volume} {545}},\ \bibinfo {pages} {323} (\bibinfo {year} {2017})}\BibitemShut
  {NoStop}%
\bibitem [{\citenamefont {Horváth}\ \emph {et~al.}(2019)\citenamefont {Horváth}, \citenamefont {Lovas}, \citenamefont {Kormos}, \citenamefont {Takács},\ and\ \citenamefont {Zaránd}}]{horvath_nonequilibrium_2019}%
  \BibitemOpen
  \bibfield  {author} {\bibinfo {author} {\bibfnamefont {D.~X.}\ \bibnamefont {Horváth}}, \bibinfo {author} {\bibfnamefont {I.}~\bibnamefont {Lovas}}, \bibinfo {author} {\bibfnamefont {M.}~\bibnamefont {Kormos}}, \bibinfo {author} {\bibfnamefont {G.}~\bibnamefont {Takács}},\ and\ \bibinfo {author} {\bibfnamefont {G.}~\bibnamefont {Zaránd}},\ }\bibfield  {title} {\bibinfo {title} {Nonequilibrium time evolution and rephasing in the quantum sine-{Gordon} model},\ }\href {https://doi.org/10.1103/PhysRevA.100.013613} {\bibfield  {journal} {\bibinfo  {journal} {Physical Review A}\ }\textbf {\bibinfo {volume} {100}},\ \bibinfo {pages} {013613} (\bibinfo {year} {2019})}\BibitemShut {NoStop}%
\bibitem [{\citenamefont {Tononi}\ \emph {et~al.}(2020)\citenamefont {Tononi}, \citenamefont {Toigo}, \citenamefont {Wimberger}, \citenamefont {Cappellaro},\ and\ \citenamefont {Salasnich}}]{tononi_dephasing_2020}%
  \BibitemOpen
  \bibfield  {author} {\bibinfo {author} {\bibfnamefont {A.}~\bibnamefont {Tononi}}, \bibinfo {author} {\bibfnamefont {F.}~\bibnamefont {Toigo}}, \bibinfo {author} {\bibfnamefont {S.}~\bibnamefont {Wimberger}}, \bibinfo {author} {\bibfnamefont {A.}~\bibnamefont {Cappellaro}},\ and\ \bibinfo {author} {\bibfnamefont {L.}~\bibnamefont {Salasnich}},\ }\bibfield  {title} {{\selectlanguage {english}\bibinfo {title} {Dephasing–rephasing dynamics of one-dimensional tunneling quasicondensates}},\ }\href {https://doi.org/10.1088/1367-2630/ab965d} {\bibfield  {journal} {\bibinfo  {journal} {New Journal of Physics}\ }\textbf {\bibinfo {volume} {22}},\ \bibinfo {pages} {073020} (\bibinfo {year} {2020})}\BibitemShut {NoStop}%
\bibitem [{\citenamefont {Bastianello}(2024)}]{bastianello_sinegordon_2024}%
  \BibitemOpen
  \bibfield  {author} {\bibinfo {author} {\bibfnamefont {A.}~\bibnamefont {Bastianello}},\ }\bibfield  {title} {\bibinfo {title} {Sine-{Gordon} model from coupled condensates: {A} generalized hydrodynamics viewpoint},\ }\href {https://doi.org/10.1103/PhysRevB.109.035118} {\bibfield  {journal} {\bibinfo  {journal} {Physical Review B}\ }\textbf {\bibinfo {volume} {109}},\ \bibinfo {pages} {035118} (\bibinfo {year} {2024})}\BibitemShut {NoStop}%
\bibitem [{\citenamefont {Szász-Schagrin}\ \emph {et~al.}(2024)\citenamefont {Szász-Schagrin}, \citenamefont {Lovas},\ and\ \citenamefont {Takács}}]{szasz-schagrin_nonequilibrium_2024}%
  \BibitemOpen
  \bibfield  {author} {\bibinfo {author} {\bibfnamefont {D.}~\bibnamefont {Szász-Schagrin}}, \bibinfo {author} {\bibfnamefont {I.}~\bibnamefont {Lovas}},\ and\ \bibinfo {author} {\bibfnamefont {G.}~\bibnamefont {Takács}},\ }\bibfield  {title} {\bibinfo {title} {Nonequilibrium time evolution in the sine-{Gordon} model},\ }\href {https://doi.org/10.1103/PhysRevB.109.014308} {\bibfield  {journal} {\bibinfo  {journal} {Physical Review B}\ }\textbf {\bibinfo {volume} {109}},\ \bibinfo {pages} {014308} (\bibinfo {year} {2024})}\BibitemShut {NoStop}%
\bibitem [{\citenamefont {Heinen}\ \emph {et~al.}(2023)\citenamefont {Heinen}, \citenamefont {Mikheev},\ and\ \citenamefont {Gasenzer}}]{heinen_anomalous_2023}%
  \BibitemOpen
  \bibfield  {author} {\bibinfo {author} {\bibfnamefont {P.}~\bibnamefont {Heinen}}, \bibinfo {author} {\bibfnamefont {A.~N.}\ \bibnamefont {Mikheev}},\ and\ \bibinfo {author} {\bibfnamefont {T.}~\bibnamefont {Gasenzer}},\ }\bibfield  {title} {\bibinfo {title} {Anomalous scaling at nonthermal fixed points of the sine-{Gordon} model},\ }\href {https://doi.org/10.1103/PhysRevA.107.043303} {\bibfield  {journal} {\bibinfo  {journal} {Physical Review A}\ }\textbf {\bibinfo {volume} {107}},\ \bibinfo {pages} {043303} (\bibinfo {year} {2023})}\BibitemShut {NoStop}%
\bibitem [{\citenamefont {Barker}\ \emph {et~al.}(2020)\citenamefont {Barker}, \citenamefont {Sunami}, \citenamefont {Garrick}, \citenamefont {Beregi}, \citenamefont {Luksch}, \citenamefont {Bentine},\ and\ \citenamefont {Foot}}]{barker_coherent_2020}%
  \BibitemOpen
  \bibfield  {author} {\bibinfo {author} {\bibfnamefont {A.~J.}\ \bibnamefont {Barker}}, \bibinfo {author} {\bibfnamefont {S.}~\bibnamefont {Sunami}}, \bibinfo {author} {\bibfnamefont {D.}~\bibnamefont {Garrick}}, \bibinfo {author} {\bibfnamefont {A.}~\bibnamefont {Beregi}}, \bibinfo {author} {\bibfnamefont {K.}~\bibnamefont {Luksch}}, \bibinfo {author} {\bibfnamefont {E.}~\bibnamefont {Bentine}},\ and\ \bibinfo {author} {\bibfnamefont {C.~J.}\ \bibnamefont {Foot}},\ }\bibfield  {title} {{\selectlanguage {english}\bibinfo {title} {Coherent splitting of two-dimensional {Bose} gases in magnetic potentials}},\ }\href {https://doi.org/10.1088/1367-2630/abbced} {\bibfield  {journal} {\bibinfo  {journal} {New Journal of Physics}\ }\textbf {\bibinfo {volume} {22}},\ \bibinfo {pages} {103040} (\bibinfo {year} {2020})}\BibitemShut {NoStop}%
\bibitem [{\citenamefont {Sunami}\ \emph {et~al.}(2022)\citenamefont {Sunami}, \citenamefont {Singh}, \citenamefont {Garrick}, \citenamefont {Beregi}, \citenamefont {Barker}, \citenamefont {Luksch}, \citenamefont {Bentine}, \citenamefont {Mathey},\ and\ \citenamefont {Foot}}]{sunami_observation_2022}%
  \BibitemOpen
  \bibfield  {author} {\bibinfo {author} {\bibfnamefont {S.}~\bibnamefont {Sunami}}, \bibinfo {author} {\bibfnamefont {V.}~\bibnamefont {Singh}}, \bibinfo {author} {\bibfnamefont {D.}~\bibnamefont {Garrick}}, \bibinfo {author} {\bibfnamefont {A.}~\bibnamefont {Beregi}}, \bibinfo {author} {\bibfnamefont {A.}~\bibnamefont {Barker}}, \bibinfo {author} {\bibfnamefont {K.}~\bibnamefont {Luksch}}, \bibinfo {author} {\bibfnamefont {E.}~\bibnamefont {Bentine}}, \bibinfo {author} {\bibfnamefont {L.}~\bibnamefont {Mathey}},\ and\ \bibinfo {author} {\bibfnamefont {C.}~\bibnamefont {Foot}},\ }\bibfield  {title} {\bibinfo {title} {Observation of the {Berezinskii}-{Kosterlitz}-{Thouless} {Transition} in a {Two}-{Dimensional} {Bose} {Gas} via {Matter}-{Wave} {Interferometry}},\ }\href {https://doi.org/10.1103/PhysRevLett.128.250402} {\bibfield  {journal} {\bibinfo  {journal} {Physical Review Letters}\ }\textbf {\bibinfo {volume} {128}},\ \bibinfo {pages} {250402} (\bibinfo {year} {2022})}\BibitemShut {NoStop}%
\bibitem [{\citenamefont {Beregi}\ \emph {et~al.}(2024)\citenamefont {Beregi}, \citenamefont {Foot},\ and\ \citenamefont {Sunami}}]{beregi_quantum_2024}%
  \BibitemOpen
  \bibfield  {author} {\bibinfo {author} {\bibfnamefont {A.}~\bibnamefont {Beregi}}, \bibinfo {author} {\bibfnamefont {C.}~\bibnamefont {Foot}},\ and\ \bibinfo {author} {\bibfnamefont {S.}~\bibnamefont {Sunami}},\ }\bibfield  {title} {\bibinfo {title} {Quantum simulations with bilayer {2D} {Bose} gases in multiple-{RF}-dressed potentials},\ }\href {https://doi.org/10.1116/5.0210068} {\bibfield  {journal} {\bibinfo  {journal} {AVS Quantum Science}\ }\textbf {\bibinfo {volume} {6}},\ \bibinfo {pages} {030501} (\bibinfo {year} {2024})}\BibitemShut {NoStop}%
\bibitem [{\citenamefont {Sunami}\ \emph {et~al.}(2025)\citenamefont {Sunami}, \citenamefont {Singh}, \citenamefont {Rydow}, \citenamefont {Beregi}, \citenamefont {Chang}, \citenamefont {Mathey},\ and\ \citenamefont {Foot}}]{sunami_detecting_2025}%
  \BibitemOpen
  \bibfield  {author} {\bibinfo {author} {\bibfnamefont {S.}~\bibnamefont {Sunami}}, \bibinfo {author} {\bibfnamefont {V.~P.}\ \bibnamefont {Singh}}, \bibinfo {author} {\bibfnamefont {E.}~\bibnamefont {Rydow}}, \bibinfo {author} {\bibfnamefont {A.}~\bibnamefont {Beregi}}, \bibinfo {author} {\bibfnamefont {E.}~\bibnamefont {Chang}}, \bibinfo {author} {\bibfnamefont {L.}~\bibnamefont {Mathey}},\ and\ \bibinfo {author} {\bibfnamefont {C.~J.}\ \bibnamefont {Foot}},\ }\bibfield  {title} {\bibinfo {title} {Detecting {Phase} {Coherence} of {2D} {Bose} {Gases} via {Noise} {Correlations}},\ }\href {https://doi.org/10.1103/PhysRevLett.134.183407} {\bibfield  {journal} {\bibinfo  {journal} {Physical Review Letters}\ }\textbf {\bibinfo {volume} {134}},\ \bibinfo {pages} {183407} (\bibinfo {year} {2025})}\BibitemShut {NoStop}%
\bibitem [{\citenamefont {Whitlock}\ and\ \citenamefont {Bouchoule}(2003)}]{whitlock_relative_2003}%
  \BibitemOpen
  \bibfield  {author} {\bibinfo {author} {\bibfnamefont {N.~K.}\ \bibnamefont {Whitlock}}\ and\ \bibinfo {author} {\bibfnamefont {I.}~\bibnamefont {Bouchoule}},\ }\bibfield  {title} {\bibinfo {title} {Relative phase fluctuations of two coupled one-dimensional condensates},\ }\href {https://doi.org/10.1103/PhysRevA.68.053609} {\bibfield  {journal} {\bibinfo  {journal} {Physical Review A}\ }\textbf {\bibinfo {volume} {68}},\ \bibinfo {pages} {053609} (\bibinfo {year} {2003})}\BibitemShut {NoStop}%
\bibitem [{\citenamefont {Grišins}\ and\ \citenamefont {Mazets}(2013)}]{grisins_coherence_2013}%
  \BibitemOpen
  \bibfield  {author} {\bibinfo {author} {\bibfnamefont {P.}~\bibnamefont {Grišins}}\ and\ \bibinfo {author} {\bibfnamefont {I.~E.}\ \bibnamefont {Mazets}},\ }\bibfield  {title} {\bibinfo {title} {Coherence and {Josephson} oscillations between two tunnel-coupled one-dimensional atomic quasicondensates at finite temperature},\ }\href {https://doi.org/10.1103/PhysRevA.87.013629} {\bibfield  {journal} {\bibinfo  {journal} {Physical Review A}\ }\textbf {\bibinfo {volume} {87}},\ \bibinfo {pages} {013629} (\bibinfo {year} {2013})}\BibitemShut {NoStop}%
\bibitem [{\citenamefont {Fertig}(2002)}]{fertig_deconfinement_2002}%
  \BibitemOpen
  \bibfield  {author} {\bibinfo {author} {\bibfnamefont {H.~A.}\ \bibnamefont {Fertig}},\ }\bibfield  {title} {\bibinfo {title} {Deconfinement in the {Two}-{Dimensional} \${\textbackslash}mathit\{{XY}\}\$ {Model}},\ }\href {https://doi.org/10.1103/PhysRevLett.89.035703} {\bibfield  {journal} {\bibinfo  {journal} {Physical Review Letters}\ }\textbf {\bibinfo {volume} {89}},\ \bibinfo {pages} {035703} (\bibinfo {year} {2002})}\BibitemShut {NoStop}%
\bibitem [{\citenamefont {Fertig}\ and\ \citenamefont {Majumdar}(2003)}]{fertig_vortex_2003}%
  \BibitemOpen
  \bibfield  {author} {\bibinfo {author} {\bibfnamefont {H.~A.}\ \bibnamefont {Fertig}}\ and\ \bibinfo {author} {\bibfnamefont {K.}~\bibnamefont {Majumdar}},\ }\bibfield  {title} {\bibinfo {title} {Vortex deconfinement in the \textit{{XY}} model with a magnetic field},\ }\href {https://doi.org/10.1016/S0003-4916(03)00061-7} {\bibfield  {journal} {\bibinfo  {journal} {Annals of Physics}\ }\textbf {\bibinfo {volume} {305}},\ \bibinfo {pages} {190} (\bibinfo {year} {2003})}\BibitemShut {NoStop}%
\bibitem [{\citenamefont {Jelić}\ and\ \citenamefont {Cugliandolo}(2011)}]{jelic_quench_2011}%
  \BibitemOpen
  \bibfield  {author} {\bibinfo {author} {\bibfnamefont {A.}~\bibnamefont {Jelić}}\ and\ \bibinfo {author} {\bibfnamefont {L.~F.}\ \bibnamefont {Cugliandolo}},\ }\bibfield  {title} {{\selectlanguage {english}\bibinfo {title} {Quench dynamics of the 2d {XY} model}},\ }\href {https://doi.org/10.1088/1742-5468/2011/02/P02032} {\bibfield  {journal} {\bibinfo  {journal} {Journal of Statistical Mechanics: Theory and Experiment}\ }\textbf {\bibinfo {volume} {2011}},\ \bibinfo {pages} {P02032} (\bibinfo {year} {2011})}\BibitemShut {NoStop}%
\bibitem [{\citenamefont {Nam}\ \emph {et~al.}(2012)\citenamefont {Nam}, \citenamefont {Baek}, \citenamefont {Kim},\ and\ \citenamefont {Lee}}]{nam_coarsening_2012}%
  \BibitemOpen
  \bibfield  {author} {\bibinfo {author} {\bibfnamefont {K.}~\bibnamefont {Nam}}, \bibinfo {author} {\bibfnamefont {W.-B.}\ \bibnamefont {Baek}}, \bibinfo {author} {\bibfnamefont {B.}~\bibnamefont {Kim}},\ and\ \bibinfo {author} {\bibfnamefont {S.~J.}\ \bibnamefont {Lee}},\ }\bibfield  {title} {{\selectlanguage {english}\bibinfo {title} {Coarsening of two-dimensional {XY} model with {Hamiltonian} dynamics: logarithmically divergent vortex mobility}},\ }\href {https://doi.org/10.1088/1742-5468/2012/11/P11023} {\bibfield  {journal} {\bibinfo  {journal} {Journal of Statistical Mechanics: Theory and Experiment}\ }\textbf {\bibinfo {volume} {2012}},\ \bibinfo {pages} {P11023} (\bibinfo {year} {2012})}\BibitemShut {NoStop}%
\bibitem [{\citenamefont {Mathey}\ and\ \citenamefont {Polkovnikov}(2010)}]{mathey_light_2010}%
  \BibitemOpen
  \bibfield  {author} {\bibinfo {author} {\bibfnamefont {L.}~\bibnamefont {Mathey}}\ and\ \bibinfo {author} {\bibfnamefont {A.}~\bibnamefont {Polkovnikov}},\ }\bibfield  {title} {\bibinfo {title} {Light cone dynamics and reverse {Kibble}-{Zurek} mechanism in two-dimensional superfluids following a quantum quench},\ }\href {https://doi.org/10.1103/PhysRevA.81.033605} {\bibfield  {journal} {\bibinfo  {journal} {Physical Review A}\ }\textbf {\bibinfo {volume} {81}},\ \bibinfo {pages} {033605} (\bibinfo {year} {2010})}\BibitemShut {NoStop}%
\bibitem [{\citenamefont {Sunami}\ \emph {et~al.}(2023)\citenamefont {Sunami}, \citenamefont {Singh}, \citenamefont {Garrick}, \citenamefont {Beregi}, \citenamefont {Barker}, \citenamefont {Luksch}, \citenamefont {Bentine}, \citenamefont {Mathey},\ and\ \citenamefont {Foot}}]{sunami_universal_2023}%
  \BibitemOpen
  \bibfield  {author} {\bibinfo {author} {\bibfnamefont {S.}~\bibnamefont {Sunami}}, \bibinfo {author} {\bibfnamefont {V.~P.}\ \bibnamefont {Singh}}, \bibinfo {author} {\bibfnamefont {D.}~\bibnamefont {Garrick}}, \bibinfo {author} {\bibfnamefont {A.}~\bibnamefont {Beregi}}, \bibinfo {author} {\bibfnamefont {A.~J.}\ \bibnamefont {Barker}}, \bibinfo {author} {\bibfnamefont {K.}~\bibnamefont {Luksch}}, \bibinfo {author} {\bibfnamefont {E.}~\bibnamefont {Bentine}}, \bibinfo {author} {\bibfnamefont {L.}~\bibnamefont {Mathey}},\ and\ \bibinfo {author} {\bibfnamefont {C.~J.}\ \bibnamefont {Foot}},\ }\bibfield  {title} {\bibinfo {title} {Universal scaling of the dynamic {BKT} transition in quenched {2D} {Bose} gases},\ }\href {https://doi.org/10.1126/science.abq6753} {\bibfield  {journal} {\bibinfo  {journal} {Science}\ }\textbf {\bibinfo {volume} {382}},\ \bibinfo {pages} {443} (\bibinfo {year} {2023})}\BibitemShut {NoStop}%
\bibitem [{\citenamefont {Berrada}\ \emph {et~al.}(2013)\citenamefont {Berrada}, \citenamefont {van Frank}, \citenamefont {Bücker}, \citenamefont {Schumm}, \citenamefont {Schaff},\ and\ \citenamefont {Schmiedmayer}}]{berrada_integrated_2013}%
  \BibitemOpen
  \bibfield  {author} {\bibinfo {author} {\bibfnamefont {T.}~\bibnamefont {Berrada}}, \bibinfo {author} {\bibfnamefont {S.}~\bibnamefont {van Frank}}, \bibinfo {author} {\bibfnamefont {R.}~\bibnamefont {Bücker}}, \bibinfo {author} {\bibfnamefont {T.}~\bibnamefont {Schumm}}, \bibinfo {author} {\bibfnamefont {J.-F.}\ \bibnamefont {Schaff}},\ and\ \bibinfo {author} {\bibfnamefont {J.}~\bibnamefont {Schmiedmayer}},\ }\bibfield  {title} {{\selectlanguage {english}\bibinfo {title} {Integrated {Mach}–{Zehnder} interferometer for {Bose}–{Einstein} condensates}},\ }\href {https://doi.org/10.1038/ncomms3077} {\bibfield  {journal} {\bibinfo  {journal} {Nature Communications}\ }\textbf {\bibinfo {volume} {4}},\ \bibinfo {pages} {2077} (\bibinfo {year} {2013})}\BibitemShut {NoStop}%
\bibitem [{\citenamefont {Spagnolli}\ \emph {et~al.}(2017)\citenamefont {Spagnolli}, \citenamefont {Semeghini}, \citenamefont {Masi}, \citenamefont {Ferioli}, \citenamefont {Trenkwalder}, \citenamefont {Coop}, \citenamefont {Landini}, \citenamefont {Pezzè}, \citenamefont {Modugno}, \citenamefont {Inguscio}, \citenamefont {Smerzi},\ and\ \citenamefont {Fattori}}]{spagnolli_crossing_2017}%
  \BibitemOpen
  \bibfield  {author} {\bibinfo {author} {\bibfnamefont {G.}~\bibnamefont {Spagnolli}}, \bibinfo {author} {\bibfnamefont {G.}~\bibnamefont {Semeghini}}, \bibinfo {author} {\bibfnamefont {L.}~\bibnamefont {Masi}}, \bibinfo {author} {\bibfnamefont {G.}~\bibnamefont {Ferioli}}, \bibinfo {author} {\bibfnamefont {A.}~\bibnamefont {Trenkwalder}}, \bibinfo {author} {\bibfnamefont {S.}~\bibnamefont {Coop}}, \bibinfo {author} {\bibfnamefont {M.}~\bibnamefont {Landini}}, \bibinfo {author} {\bibfnamefont {L.}~\bibnamefont {Pezzè}}, \bibinfo {author} {\bibfnamefont {G.}~\bibnamefont {Modugno}}, \bibinfo {author} {\bibfnamefont {M.}~\bibnamefont {Inguscio}}, \bibinfo {author} {\bibfnamefont {A.}~\bibnamefont {Smerzi}},\ and\ \bibinfo {author} {\bibfnamefont {M.}~\bibnamefont {Fattori}},\ }\bibfield  {title} {\bibinfo {title} {Crossing {Over} from {Attractive} to {Repulsive} {Interactions} in a {Tunneling} {Bosonic} {Josephson} {Junction}},\ }\href {https://doi.org/10.1103/PhysRevLett.118.230403} {\bibfield  {journal}
  {\bibinfo  {journal} {Physical Review Letters}\ }\textbf {\bibinfo {volume} {118}},\ \bibinfo {pages} {230403} (\bibinfo {year} {2017})}\BibitemShut {NoStop}%
\bibitem [{\citenamefont {Levy}\ \emph {et~al.}(2007)\citenamefont {Levy}, \citenamefont {Lahoud}, \citenamefont {Shomroni},\ and\ \citenamefont {Steinhauer}}]{levy_ac_2007}%
  \BibitemOpen
  \bibfield  {author} {\bibinfo {author} {\bibfnamefont {S.}~\bibnamefont {Levy}}, \bibinfo {author} {\bibfnamefont {E.}~\bibnamefont {Lahoud}}, \bibinfo {author} {\bibfnamefont {I.}~\bibnamefont {Shomroni}},\ and\ \bibinfo {author} {\bibfnamefont {J.}~\bibnamefont {Steinhauer}},\ }\bibfield  {title} {{\selectlanguage {english}\bibinfo {title} {The a.c. and d.c. {Josephson} effects in a {Bose}–{Einstein} condensate}},\ }\href {https://doi.org/10.1038/nature06186} {\bibfield  {journal} {\bibinfo  {journal} {Nature}\ }\textbf {\bibinfo {volume} {449}},\ \bibinfo {pages} {579} (\bibinfo {year} {2007})}\BibitemShut {NoStop}%
\bibitem [{\citenamefont {Pigneur}\ \emph {et~al.}(2018)\citenamefont {Pigneur}, \citenamefont {Berrada}, \citenamefont {Bonneau}, \citenamefont {Schumm}, \citenamefont {Demler},\ and\ \citenamefont {Schmiedmayer}}]{pigneur_relaxation_2018}%
  \BibitemOpen
  \bibfield  {author} {\bibinfo {author} {\bibfnamefont {M.}~\bibnamefont {Pigneur}}, \bibinfo {author} {\bibfnamefont {T.}~\bibnamefont {Berrada}}, \bibinfo {author} {\bibfnamefont {M.}~\bibnamefont {Bonneau}}, \bibinfo {author} {\bibfnamefont {T.}~\bibnamefont {Schumm}}, \bibinfo {author} {\bibfnamefont {E.}~\bibnamefont {Demler}},\ and\ \bibinfo {author} {\bibfnamefont {J.}~\bibnamefont {Schmiedmayer}},\ }\bibfield  {title} {{\selectlanguage {english}\bibinfo {title} {Relaxation to a {Phase}-{Locked} {Equilibrium} {State} in a {One}-{Dimensional} {Bosonic} {Josephson} {Junction}}},\ }\href {https://doi.org/10.1103/PhysRevLett.120.173601} {\bibfield  {journal} {\bibinfo  {journal} {Physical Review Letters}\ }\textbf {\bibinfo {volume} {120}},\ \bibinfo {pages} {173601} (\bibinfo {year} {2018})}\BibitemShut {NoStop}%
\bibitem [{\citenamefont {Luick}\ \emph {et~al.}(2020)\citenamefont {Luick}, \citenamefont {Sobirey}, \citenamefont {Bohlen}, \citenamefont {Singh}, \citenamefont {Mathey}, \citenamefont {Lompe},\ and\ \citenamefont {Moritz}}]{luick_ideal_2020}%
  \BibitemOpen
  \bibfield  {author} {\bibinfo {author} {\bibfnamefont {N.}~\bibnamefont {Luick}}, \bibinfo {author} {\bibfnamefont {L.}~\bibnamefont {Sobirey}}, \bibinfo {author} {\bibfnamefont {M.}~\bibnamefont {Bohlen}}, \bibinfo {author} {\bibfnamefont {V.~P.}\ \bibnamefont {Singh}}, \bibinfo {author} {\bibfnamefont {L.}~\bibnamefont {Mathey}}, \bibinfo {author} {\bibfnamefont {T.}~\bibnamefont {Lompe}},\ and\ \bibinfo {author} {\bibfnamefont {H.}~\bibnamefont {Moritz}},\ }\bibfield  {title} {\bibinfo {title} {An ideal {Josephson} junction in an ultracold two-dimensional {Fermi} gas},\ }\href {https://doi.org/10.1126/science.aaz2342} {\bibfield  {journal} {\bibinfo  {journal} {Science}\ }\textbf {\bibinfo {volume} {369}},\ \bibinfo {pages} {89} (\bibinfo {year} {2020})}\BibitemShut {NoStop}%
\bibitem [{\citenamefont {Singh}\ and\ \citenamefont {Mathey}(2020)}]{singh_sound_2020}%
  \BibitemOpen
  \bibfield  {author} {\bibinfo {author} {\bibfnamefont {V.~P.}\ \bibnamefont {Singh}}\ and\ \bibinfo {author} {\bibfnamefont {L.}~\bibnamefont {Mathey}},\ }\bibfield  {title} {\bibinfo {title} {Sound propagation in a two-dimensional {Bose} gas across the superfluid transition},\ }\href {https://doi.org/10.1103/PhysRevResearch.2.023336} {\bibfield  {journal} {\bibinfo  {journal} {Physical Review Research}\ }\textbf {\bibinfo {volume} {2}},\ \bibinfo {pages} {023336} (\bibinfo {year} {2020})}\BibitemShut {NoStop}%
\bibitem [{\citenamefont {Singh}\ and\ \citenamefont {Mathey}(2022)}]{singh_first_2022}%
  \BibitemOpen
  \bibfield  {author} {\bibinfo {author} {\bibfnamefont {V.~P.}\ \bibnamefont {Singh}}\ and\ \bibinfo {author} {\bibfnamefont {L.}~\bibnamefont {Mathey}},\ }\bibfield  {title} {{\selectlanguage {english}\bibinfo {title} {First and second sound in a dilute {Bose} gas across the {BKT} transition}},\ }\href {https://doi.org/10.1088/1367-2630/ac7d6f} {\bibfield  {journal} {\bibinfo  {journal} {New Journal of Physics}\ }\textbf {\bibinfo {volume} {24}},\ \bibinfo {pages} {073024} (\bibinfo {year} {2022})}\BibitemShut {NoStop}%
\bibitem [{\citenamefont {Mora}\ and\ \citenamefont {Castin}(2003)}]{mora_extension_2003}%
  \BibitemOpen
  \bibfield  {author} {\bibinfo {author} {\bibfnamefont {C.}~\bibnamefont {Mora}}\ and\ \bibinfo {author} {\bibfnamefont {Y.}~\bibnamefont {Castin}},\ }\bibfield  {title} {\bibinfo {title} {Extension of {Bogoliubov} theory to quasicondensates},\ }\href {https://doi.org/10.1103/PhysRevA.67.053615} {\bibfield  {journal} {\bibinfo  {journal} {Physical Review A}\ }\textbf {\bibinfo {volume} {67}},\ \bibinfo {pages} {053615} (\bibinfo {year} {2003})}\BibitemShut {NoStop}%
\bibitem [{\citenamefont {Ananikian}\ and\ \citenamefont {Bergeman}(2006)}]{ananikian_grosspitaevskii_2006}%
  \BibitemOpen
  \bibfield  {author} {\bibinfo {author} {\bibfnamefont {D.}~\bibnamefont {Ananikian}}\ and\ \bibinfo {author} {\bibfnamefont {T.}~\bibnamefont {Bergeman}},\ }\bibfield  {title} {\bibinfo {title} {Gross-{Pitaevskii} equation for {Bose} particles in a double-well potential: {Two}-mode models and beyond},\ }\href {https://doi.org/10.1103/PhysRevA.73.013604} {\bibfield  {journal} {\bibinfo  {journal} {Physical Review A: Atomic, Molecular, and Optical Physics}\ }\textbf {\bibinfo {volume} {73}},\ \bibinfo {pages} {013604} (\bibinfo {year} {2006})}\BibitemShut {NoStop}%
\bibitem [{\citenamefont {van Nieuwkerk}\ and\ \citenamefont {Essler}(2020)}]{vannieuwkerk_lowenergy_2020}%
  \BibitemOpen
  \bibfield  {author} {\bibinfo {author} {\bibfnamefont {Y.~D.}\ \bibnamefont {van Nieuwkerk}}\ and\ \bibinfo {author} {\bibfnamefont {F.}~\bibnamefont {Essler}},\ }\bibfield  {title} {{\selectlanguage {english}\bibinfo {title} {On the low-energy description for tunnel-coupled one-dimensional {Bose} gases}},\ }\href {https://doi.org/10.21468/SciPostPhys.9.2.025} {\bibfield  {journal} {\bibinfo  {journal} {SciPost Physics}\ }\textbf {\bibinfo {volume} {9}},\ \bibinfo {pages} {025} (\bibinfo {year} {2020})}\BibitemShut {NoStop}%
\bibitem [{\citenamefont {Harte}\ \emph {et~al.}(2018)\citenamefont {Harte}, \citenamefont {Bentine}, \citenamefont {Luksch}, \citenamefont {Barker}, \citenamefont {Trypogeorgos}, \citenamefont {Yuen},\ and\ \citenamefont {Foot}}]{harte_ultracold_2018}%
  \BibitemOpen
  \bibfield  {author} {\bibinfo {author} {\bibfnamefont {T.~L.}\ \bibnamefont {Harte}}, \bibinfo {author} {\bibfnamefont {E.}~\bibnamefont {Bentine}}, \bibinfo {author} {\bibfnamefont {K.}~\bibnamefont {Luksch}}, \bibinfo {author} {\bibfnamefont {A.~J.}\ \bibnamefont {Barker}}, \bibinfo {author} {\bibfnamefont {D.}~\bibnamefont {Trypogeorgos}}, \bibinfo {author} {\bibfnamefont {B.}~\bibnamefont {Yuen}},\ and\ \bibinfo {author} {\bibfnamefont {C.~J.}\ \bibnamefont {Foot}},\ }\bibfield  {title} {\bibinfo {title} {Ultracold atoms in multiple radio-frequency dressed adiabatic potentials},\ }\href {https://doi.org/10.1103/PhysRevA.97.013616} {\bibfield  {journal} {\bibinfo  {journal} {Physical Review A}\ }\textbf {\bibinfo {volume} {97}},\ \bibinfo {pages} {013616} (\bibinfo {year} {2018})}\BibitemShut {NoStop}%
\bibitem [{\citenamefont {Bentine}\ \emph {et~al.}(2020)\citenamefont {Bentine}, \citenamefont {Barker}, \citenamefont {Luksch}, \citenamefont {Sunami}, \citenamefont {Harte}, \citenamefont {Yuen}, \citenamefont {Foot}, \citenamefont {Owens},\ and\ \citenamefont {Hutson}}]{bentine_2020}%
  \BibitemOpen
  \bibfield  {author} {\bibinfo {author} {\bibfnamefont {E.}~\bibnamefont {Bentine}}, \bibinfo {author} {\bibfnamefont {A.~J.}\ \bibnamefont {Barker}}, \bibinfo {author} {\bibfnamefont {K.}~\bibnamefont {Luksch}}, \bibinfo {author} {\bibfnamefont {S.}~\bibnamefont {Sunami}}, \bibinfo {author} {\bibfnamefont {T.~L.}\ \bibnamefont {Harte}}, \bibinfo {author} {\bibfnamefont {B.}~\bibnamefont {Yuen}}, \bibinfo {author} {\bibfnamefont {C.~J.}\ \bibnamefont {Foot}}, \bibinfo {author} {\bibfnamefont {D.~J.}\ \bibnamefont {Owens}},\ and\ \bibinfo {author} {\bibfnamefont {J.~M.}\ \bibnamefont {Hutson}},\ }\bibfield  {title} {\bibinfo {title} {Inelastic collisions in radiofrequency-dressed mixtures of ultracold atoms},\ }\href {https://doi.org/10.1103/PhysRevResearch.2.033163} {\bibfield  {journal} {\bibinfo  {journal} {Phys. Rev. Res.}\ }\textbf {\bibinfo {volume} {2}},\ \bibinfo {pages} {033163} (\bibinfo {year} {2020})}\BibitemShut {NoStop}%
\bibitem [{\citenamefont {Pethick}\ and\ \citenamefont {Smith}(2008)}]{pethick_bose_2008}%
  \BibitemOpen
  \bibfield  {author} {\bibinfo {author} {\bibfnamefont {C.~J.}\ \bibnamefont {Pethick}}\ and\ \bibinfo {author} {\bibfnamefont {H.}~\bibnamefont {Smith}},\ }\href {https://www.cambridge.org/core/books/boseeinstein-condensation-in-dilute-gases/CC439EAD70D78E47E9AF536DA7B203EC} {\emph {\bibinfo {title} {Bose–{Einstein} {Condensation} in {Dilute} {Gases}}}},\ \bibinfo {edition} {2nd}\ ed.\ (\bibinfo  {publisher} {Cambridge University Press},\ \bibinfo {address} {Cambridge},\ \bibinfo {year} {2008})\BibitemShut {NoStop}%
\bibitem [{\citenamefont {Hadzibabic}\ \emph {et~al.}(2006)\citenamefont {Hadzibabic}, \citenamefont {Krüger}, \citenamefont {Cheneau}, \citenamefont {Battelier},\ and\ \citenamefont {Dalibard}}]{hadzibabic_berezinskii_2006}%
  \BibitemOpen
  \bibfield  {author} {\bibinfo {author} {\bibfnamefont {Z.}~\bibnamefont {Hadzibabic}}, \bibinfo {author} {\bibfnamefont {P.}~\bibnamefont {Krüger}}, \bibinfo {author} {\bibfnamefont {M.}~\bibnamefont {Cheneau}}, \bibinfo {author} {\bibfnamefont {B.}~\bibnamefont {Battelier}},\ and\ \bibinfo {author} {\bibfnamefont {J.}~\bibnamefont {Dalibard}},\ }\bibfield  {title} {{\selectlanguage {english}\bibinfo {title} {Berezinskii–{Kosterlitz}–{Thouless} crossover in a trapped atomic gas}},\ }\href {https://doi.org/10.1038/nature04851} {\bibfield  {journal} {\bibinfo  {journal} {Nature}\ }\textbf {\bibinfo {volume} {441}},\ \bibinfo {pages} {1118} (\bibinfo {year} {2006})}\BibitemShut {NoStop}%
\bibitem [{\citenamefont {Beregi}(2024)}]{beregi_thesis}%
  \BibitemOpen
  \bibfield  {author} {\bibinfo {author} {\bibfnamefont {A.}~\bibnamefont {Beregi}},\ }\emph {\bibinfo {title} {Probing universality of 2D quantum systems with bilayer Bose gases}},\ \href@noop {} {Ph.D. thesis},\ \bibinfo  {school} {University of Oxford} (\bibinfo {year} {2024})\BibitemShut {NoStop}%
\bibitem [{\citenamefont {Bayocboc}\ \emph {et~al.}(2022)\citenamefont {Bayocboc}, \citenamefont {Davis},\ and\ \citenamefont {Kheruntsyan}}]{bayocboc_dynamics_2022}%
  \BibitemOpen
  \bibfield  {author} {\bibinfo {author} {\bibfnamefont {F.~A.}\ \bibnamefont {Bayocboc}}, \bibinfo {author} {\bibfnamefont {M.~J.}\ \bibnamefont {Davis}},\ and\ \bibinfo {author} {\bibfnamefont {K.~V.}\ \bibnamefont {Kheruntsyan}},\ }\bibfield  {title} {\bibinfo {title} {Dynamics of thermalization of two tunnel-coupled one-dimensional quasicondensates},\ }\href {https://doi.org/10.1103/PhysRevA.106.023320} {\bibfield  {journal} {\bibinfo  {journal} {Physical Review A}\ }\textbf {\bibinfo {volume} {106}},\ \bibinfo {pages} {023320} (\bibinfo {year} {2022})}\BibitemShut {NoStop}%
\end{thebibliography}
\end{document}